\documentclass[preprints,review,accept,pdftex,moreauthors]{mdpi} 
\usepackage{upgreek}

\firstpage{1} 
\pubvolume{1}
\issuenum{1}
\articlenumber{0}
\pubyear{2023}
\copyrightyear{2023}
\datereceived{8 April 2023 } 
\daterevised{29 June 2023 } 
\dateaccepted{7 July 2023 } 
\datepublished{ } 
\hreflink{https://doi.org/} 

\usepackage{comment}
\usepackage{astrojournals}

\Title{Multidimensional Simulations of Core Convection}

\TitleCitation{Multidimensional Simulations of Core Convection}

\Author{Daniel Lecoanet $^{1,2,}$*
$^{,\dagger}$\orcidA{} and Philipp V. F. Edelmann $^{3,4,}$*$^{,\dagger}$\orcidB{}}

\AuthorNames{Daniel Lecoanet  and Philipp V. F. Edelmann}

\AuthorCitation{Lecoanet, D.; Edelmann, P.V.F.}

\address{%
$^{1}$ \quad CIERA, Northwestern University, Evanston, IL 60201, USA\\
$^{2}$ \quad Department of Engineering Sciences and Applied Mathematics, Northwestern University, \mbox{Evanston, IL 60208, USA}\\
$^{3}$ \quad Computer, Computational and Statistical Sciences (CCS) Division, Los Alamos National Laboratory, Los Alamos, NM 87545, USA\\
$^{4}$ \quad Center for Theoretical Astrophysics (CTA), Los Alamos National Laboratory, Los Alamos, NM 87545, USA}

\corres{Correspondence: daniel.lecoanet@northwestern.edu (D.L.); pedelmann@lanl.gov (P.V.F.E.)}

\firstnote{These authors contributed equally to this work.}

\abstract{
The cores of main sequence intermediate- and high-mass stars are convective.
Mixing at the radiative--convective boundary, waves excited by the convection, and magnetic fields generated by convective dynamos all influence the main sequence and post-main sequence evolution of these stars.
These effects must be understood to accurately model the structure and evolution of intermediate- and high-mass stars.
Unfortunately, there are many challenges in simulating core convection due to the wide range of temporal and spatial scales, as well as many important physics effects.
In this review, we describe the latest numerical strategies to address these challenges.
We then describe the latest state-of-the-art simulations of core convection, summarizing their main findings.
These simulations have led to important insights into many of the processes associated with core convection.
Two outstanding problems with multidimensional simulations are, 1.~it is not always straightforward to extrapolate from simulation parameters to the parameters of real stars; and 2.~simulations using different methods sometimes appear to arrive at contradictory results.
To address these issues, next generation simulations of core convection must address how their results depend on stellar luminosity, dimensionality, and turbulence intensity.
Furthermore, code comparison projects will be essential to establish robust parameterizations that will become the new standard in stellar modeling.}

\keyword{magnetohydrodynamics; convection zones; internal waves; computational methods}

\newcommand{\PD}[2]{\frac{\partial#1}{\partial#2}}

\newcommand{\newtext}[1]{ #1 }

\begin{document}

 \section{Introduction and Motivation}

All \textls[-25]{main-sequence stars have convection zones; while lower-mass stars with \mbox{$M\lesssim 1.5 M_{\odot}$}} have convective envelopes, higher-mass stars with $M\gtrsim 1.5 M_{\odot}$ have convective cores.
These higher-mass stars with convective cores are rare, but disproportionately important in astrophysics.
Massive stars ($M\gtrsim 8_{\odot}$) are the progenitors of neutron stars and black holes, and both massive stars and intermediate-mass stars ($1.5M_\odot\lesssim M\lesssim8 M_{\odot}$) chemically enrich their environments through winds and/or explosive mass-loss.
The lives and deaths of intermediate-mass and massive stars are influenced by the convective processes in their cores while they are on the main sequence.

Despite their importance, there remain significant uncertainties in the main-sequence lifetime of stars with convective cores (e.g., \citep{Castro2014}).
While the convection zone is chemically homogeneous, fresh H fuel can be mixed into the core by \textit{convective boundary mixing}, extending the star's main-sequence lifetime.
We use the term convective boundary mixing to encompass any physical mechanism which brings radiative-zone material into the convection zone  (e.g., convective overshoot and convective penetration \citet{anders2022}).
\mbox{\citet{kaiser2020}} examined a series of plausible models of convective boundary mixing, and found variations of $\sim 20$\% in the main-sequence lifetimes of stars between 15 and 20$M_{\odot}$.
These differences compound over the life of the star; they found differences in He core masses of $\sim$40\%, and differences in CO core masses of $\sim$60\%.
Realistic models of convective boundary mixing must be implemented in stellar evolution codes and population synthesis models in order to accurately predict properties of stellar remnants \citep{agrawal2022} and nucleosynthetic yields \citep{Arnett2011}.

Magnetic fields are also very important in intermediate-mass and massive stars.
Strong stellar magnetic fields may power superluminous supernova and seed $\sim$$10^{15}\,{\rm G}$ magnetar magnetic fields \citep{mosta2015}.
Torques from the Lorentz force can lead to very efficient angular momentum transport \citep{ferraro1937, gough1998}.
Although there is substantial controversy over magnetic field generation in the radiative zones of stars \citep{spruit2002, fuller2019, braithwaite2006, zahn2007, petitdemange2022}, it is well-established that convection can generate magnetic fields (e.g., \citep{moffatt2019}) The cores of intermediate-mass and massive stars likely harbor large-scale magnetic fields \citep{Augustson2016}, though the effects of these magnetic fields on convective boundary mixing, angular momentum transport, and the subsequent post-main-sequence evolution of these stars is as-yet unknown.
Recently, the presence of magnetic fields has been inferred from asteroseismic analysis in a main-sequence B star \citep{Lecoanet2022}, and RGB stars with masses $\approx 1.5 M_{\odot}$ \citep{fuller2015, li2022, Deheuvels2023}.
These magnetic fields could be generated by core convective dynamos while the stars were on the main sequence, or they may be remnants of fossil magnetic fields.

Although the most vigorous motions in intermediate-mass and massive stars occur in their convective cores, their radiative envelopes are anything but quiescent.
These radiative zones host a wide variety of waves; in particular, internal gravity waves, the propagating version of $g$-modes.
In intermediate-mass and massive stars, internal gravity waves are generated by convection, as well as by opacity gradients in the outer part of the star.
These waves can transport angular momentum \citep{Rogers2013,Talon2003,Townsend2018}, as well as chemicals (e.g., \citep{Rogers2017}) in stellar radiative envelopes.
The latest asteroseismic inferences of chemical mixing in stars may be able to probe mixing by waves \citep{Pedersen2021}.
Furthermore, convectively excited waves may be visible at the stellar surface (e.g., \citep{Aerts2015,horst2020a}). \citet{Bowman2019} have detected ubiquitous low-frequency variability in massive stars, which could be from convectively excited waves.
If so, this variability could encode important properties of core convection.

While many stellar processes are adequately described by one-dimensional stellar structure and evolution calculations, convection is a fundamentally multidimensional phenomenon that must be parameterized. ({Luckily the thermal structure of the convection zone is well-known; the convection is efficient so the temperature gradient is adiabatic. Thus, mixing length theory is not necessary for core convection and core convection zones are not affected by different choices or implementations for mixing length parameters.})
Convective boundary mixing, dynamo magnetic field generation, and wave-driven angular momentum and chemical transport all affect the evolution of intermediate-mass and massive stars, and must be included for high-fidelity stellar modeling, comparing to observations, and making predictions for population synthesis and nucleosynthetic models.
In this review, we discuss the current state-of-the-art multidimensional simulations of stellar core convection, and how these simulations are being used to understand the processes which govern stellar evolution.

 \section{Simulation Challenges}\label{sec:challenges}

While core convection is important in astrophysics, it is very difficult to study from first principles.
The fluid motions of core convection are governed by the Navier--Stokes equation, but are also influenced by, \newtext{e.g., magnetic fields, rotation,} nuclear reactions, and the opacity and equation of state of the constituent plasma.
There are no relevant analytic solutions to this system of algebraic and partial differential equations, so the equations must be simulated numerically.
We now encounter the fundamental challenge of astrophysical fluid dynamics, to run a direct numerical simulation of a star using current hardware would require the entire power output of an M dwarf \citep{Tobias2021}.
\newtext{Others have described some of the general challenges of studying stellar convection in solar-type stars \cite{Kupka2017}.} \newtext{Here,} we describe some of the main challenges facing numerical simulations of core convection; Section~\ref{sec:approaches} describes how different simulation codes address these challenges.

There are many important processes associated with core convection, each of which have different characteristic time and length scales.
A major challenge in simulating core convection is capturing these very discrepant time and length scales.
We will illustrate these ranges by using a zero-age main sequence $10M_\odot$ star as an example.
Although the precise values of each time and length scale is different for intermediate- and high-mass stars of different masses and ages, the order of magnitude is similar.
We use a stellar structure model from \citet{Lecoanet2019}, but different stellar modeling choices do not significantly impact the values reported here.
\citet{Jermyn2022a} describes how some of these time and length scales vary with mass and age over the main sequence.

\subsection{Time Scales}

\subsubsection{Sound}

A typical sound speed in the convective core of a $10 M_\odot$ zero-age main sequence star is $c_s\sim 7\times10^{8}\,{\rm cm}/{\rm s}$. The radius of the convection zone is $R_c\sim 6\times 10^{10}\,{\rm cm}$, so the sound-crossing time (or dynamical) across the convection zone is
\begin{equation}
t_{\rm sound} = \frac{R_c}{c_s} \sim 10^2\,{\rm s}.
\end{equation}
This estimates the time scale over which fluid motions in the convection zone reach pressure equilibrium.
This is one of the shortest global time scales of the star, meaning that the fluid is always close to pressure equilibrium.
An important practical concern is the sound-wave Courant--Friedrichs--Lewy (CFL) condition, which states that numerical simulations which use explicit timestepping methods to solve the compressible Navier--Stokes equations must take timesteps of size $\lesssim$$h/c_s$, where $h$ is the simulation grid resolution.
Thus, a high-resolution simulation with $10^3$ grid points across the convection zone can only take timesteps of size $\lesssim$$0.1\,{\rm s}$ if the simulation uses an explicit timestepping scheme.

\subsubsection{Buoyancy}

The radiative zone of a star is stably stratified and admits waves known as internal gravity waves or buoyancy waves.
The \newtext{shortest} possible period of an internal gravity wave is given by the buoyancy time
\begin{equation}
t_{\rm b} = \frac{2\pi}{N} = 2\pi \left(\frac{g}{c_p}\frac{\partial s}{\partial r}\right)^{-1/2} \sim 10^{3}\,{\rm s},
\end{equation}
where $N$ is known as the Brunt--V\"ais\"al\"a frequency or the buoyancy frequency, $g$ is the gravitational acceleration, $c_p$ is the ratio of specific heats at constant pressure, $s$ is the specific entropy, and $r$ is the radius.
The buoyancy time is typically similar to the sound time as they are both roughly $H_p/c_s$, where $H_p$ is the pressure scale height.
Note, however, that sound waves propagate on time scales shorter than $t_{\rm sound}$, whereas buoyancy waves propagate on time scales longer than $t_{\rm b}$.
The buoyancy time gives an estimate of the strength of stable stratification in the radiative zone.

\subsubsection{Convection}\label{sec:turnover}

We estimate the convective time as the ratio of the radius of the convection zone $R_c$ to the convective velocity $u_c$,
\begin{equation}
t_{\rm c} = \frac{R_c}{u_c} = R_c \left(\frac{L_c}{4\pi r^2 \rho}\right)^{-1/3}\sim 10^6\, {\rm s}.
\end{equation}
This measures the time it takes for a fluid parcel to move from the center of the star to the edge of the convection zone.
We estimate the convective velocity by
\begin{equation}
\rho u_c^3 = F_c = \frac{L_c}{4\pi r^2},
\end{equation}
where $F_c$ is the convective flux, $L_c$ is the convective luminosity, and $\rho$ is the density.
\newtext{The convective luminosity is the difference of the total luminosity and the radiative luminosity.}

Here we see that the convective time is $\sim$$10^4$ times longer than the sound-crossing time.
The former means that simulations which use explicit timestepping schemes require timesteps that are $10^4$ smaller than simulations using implicit timestepping schemes.
Such a simulation with $10^3$ grid points across the convection requires $\mathcal{O}(10^7)$ timesteps per convection time.
The convection time is also $\sim$$10^3$ times longer than the buoyancy time.
That means the radiative zone is a thousand times as stably stratified as the convection zone is unstable to convection.
Thus, convection produces only very slight motions in the radiative zone.

\subsubsection{Thermal}\label{sec:thermal-time-scale}

Stars maintain flux equilibrium on the thermal, or Kelvin--Helmholtz, time scale.
Here we estimate the thermal time scale using the radiative diffusivity $k_\mathrm{rad}$ and the radius of the convective core,
\begin{equation}
t_{\rm th} = \frac{R_c^2}{k_\mathrm{rad}} \sim 10^{12}\,{\rm s}.
\end{equation}
The radiative diffusivity is given by
\begin{equation}
k_{\rm rad} = \frac{16\sigma T^3}{3\kappa c_p \rho^2},
\end{equation}
\newtext{where $\sigma$ is the Stefan--Boltzmann constant, $T$ is the temperature, and $\kappa$ is the opacity.}
Thus, there are about one million convection times in a single thermal time.
When running a convection simulation, there is often a quick transient phase $\mathcal{O}(t_{\rm c})$ over which convection establishes itself.
While many convective properties are steady after this transient phase, there are some secular changes which occur over the thermal time.
Most significantly, the region near the boundary between the convective and radiative zone takes a thermal time to equilibrate \citep{Anders2022penetration}.

\subsubsection{Evolution}

The longest time scale relevant for core convection is the main-sequence lifetime of the star.
This is the nuclear burning time scale.
For a $10M_\odot$ star, this is about
\begin{equation}
t_{\rm e}\sim 10^{15}\,{\rm s},
\end{equation}
or about 1000 thermal times.
It is neither possible nor necessary to run multidimensional convection simulations on evolution times.
Because $t_{\rm e}\gg t_{\rm th}$, it is sufficient to solve for the thermal equilibrium of the convective system.
To capture the evolving thermal equilibrium across the main-sequence lifetime, one can either (i) perform a limited number of simulations at different stages of evolution; or (ii) parameterize the equilibrium state in terms of properties that vary across the main-sequence lifetime (e.g., ratio of buoyancy to convective time scales).

\subsection{Length Scales}

\subsubsection{Convection Zone Radius}

The radius of the convection zone of a $10M_\odot$ zero age main sequence star is about
\begin{equation}
R_c\sim 6\times 10^{10}\,{\rm cm}.
\end{equation}
Absent rotational effects, the core convection organizes itself into a dipole flow going through the center of the star and returning along the radiative--convective boundary (e.g.,~\citep{Woodward2019}).
Hence, convective flows can organize on length scales as large as $R_c$.

\subsubsection{Pressure Scale Height}

The pressure scale height
\begin{equation}
H_p = p \ \left(-\frac{d p}{dr}\right)^{-1} \sim 3\times 10^{10}\, {\rm cm}
\end{equation}
measures the length scale over which background thermodynamic quantities vary in the star.
If fluid moves across multiple density scale heights, compressibility effects can lead to important asymmetries and organizations of the flow.
However, in core convection, the pressure scale height is similar to the convection zone radius, so compressibility plays a minor role.

\subsubsection{Overshoot Length}

How far do convective motions overshoot into the radiative zone?
This is important for describing convective boundary mixing and the generation of internal gravity waves by convection.
Here, we estimate the overshoot length by equating the kinetic energy \newtext{of convective flows} $\frac{1}{2}\rho u_c^2$ to the potential energy cost for rising a fluid element a height $\ell_{\rm ov}$ in the radiative zone $\frac{1}{2}\rho \ell_{\rm ov}^2N^2$,
\begin{equation}
\ell_{\rm ov}= \frac{u_c}{N}\sim 6\times 10^{7}\, {\rm cm}.
\end{equation}

This estimate of the overshoot length is smaller than the convection zone radius by a factor of $10^3$; see \citet{Anders2023} for other ways of estimating the overshoot length.
Simulations thus require a very fine resolution near the radiative--convective boundary to accurately simulate overshooting dynamics.

\subsubsection{Radiative Diffusion Length}

Radiation is an essential energy transport mechanism in stars.
However, radiative diffusion also affects small-scale fluid motions in the convection zone.
On the radiative diffusion length scale, fluid motions will diffuse their thermal content within one \newtext{eddy} turnover time.
Motions smaller than the radiative diffusion length scale are not affected by buoyancy.
\newtext{In Section~\ref{sec:turnover}, we estimated the convective turnover time for the largest scale convective flows.
Similarly, one can define the turnover time of a small-scale eddy of size $1/k$ with velocity $u_k$ to be $\tau_\ell=(k\,u_\ell)^{-1}$.}
We will assume a Kolmogorov $k^{-5/3}$ kinetic energy spectrum, where $k$ is the wavenumber associated with convective motions.
This implies the \newtext{eddy} turnover time scales $\sim k^{-2/3}$.
Then the radiative diffusion length scale is
\begin{equation}
\ell_{\rm d} = R_c \left(\frac{u_c R_c}{k_\mathrm{rad}}\right)^{-3/4} \sim 2\times 10^{6}\,{\rm cm}.
\end{equation}

The radiative diffusion length is smaller than the overshoot length, but not significantly so.
A simulation with 3 grid points per radiative diffusion length would require $\sim$$10^5$ grid points across the convection zone.
This is beyond current computational capabilities.

\subsubsection{Viscous Length}

The viscosity of stellar plasmas is significantly smaller than the radiative diffusivity.
This mean that there are fluid motions below the radiative diffusion length, even if they are no longer driven by buoyancy.
These motions should follow a Kolmogorov kinetic energy cascade, continuously transfer energy to smaller scales until it is viscously dissipated.
Similar to the radiative diffusion length, we estimate the viscous length is
\begin{equation}
\ell_{\rm v} = R_c \left(\frac{u_c R_c}{\nu}\right)^{-3/4} \sim 300\,{\rm cm},
\end{equation}
where $\nu$ is the viscosity \citep{Jermyn2022a}.
This is the smallest important length scale for core convection; below $\ell_v$, there are negligible motions.
Because it is so much smaller than the radius of the convection zone, there is no hope to simulate all scales of motion in core convection.
That is to say, numerical simulations are less turbulent than real stars.
\newtext{Numerical simulations must introduce dissipation on much larger scales than $\ell_{\rm v}$, which is sometimes captured by an \textit{artificial viscosity}.}
Fortunately, energy in turbulence is predominately transferred from large scales to small scales, which is easy to simulate.
Nevertheless, there is also \textit{backscatter}, in which small scale flows can influence large scale flows (e.g., \citep{Jofre2018}).
This does not occur properly in numerical simulations, which do not capture the full range of scales of core convection.

\subsection{Physical Effects}

Beyond the wide range of time and length scales, there are many physical effects which are challenging to include in numerical simulations of core convection.
Here, we will include a brief description of the main challenges.

\subsubsection{Spherical Geometry}

Core convection occurs in a roughly spherical region of a star.
This makes it natural to use spherical coordinates to describe core convection.
However, the spherical coordinate system has coordinate singularities at $r=0$ and at the poles ($\theta=0, \pi$).
Different variables must satisfy different regularity conditions at the poles \citep{Vasil2019}, which can be difficult to impose.
Some simulations avoid these coordinate singularities by excising small regions near the origin and/or the poles, but these artificial surfaces can lead to extra friction and change large-scale flow characteristics.
Without rotation, core convection organizes into a large-scale dipole flow which passes directly through the origin, excising a small region around the origin can disrupt this flow.

Another strategy is to solve the problem in Cartesian geometry.
While this avoids coordinate singularities, it does not conform to the spherical symmetry of the star.
In particular, recall that the overshoot length scale is roughly 1000 times smaller than the convection zone radius.
That means the radiative--convective boundary is very nearly spherical.
Simulations require very fine grid spacing near the radiative--convective boundary so they can adequately separate convective from radiative regions.
This is easier to accomplish when using spherical geometry, as one can use small radial grid spacing near the radiative--convective boundary.
When using Cartesian coordinates, higher radial resolution near the radiative--convective boundary requires multidimensional grid~refinement.

\subsubsection{Rotation}

Intermediate-mass and massive stars rotate, which influences the convection in their cores.
Typical photometric rotation periods for B stars are about $T=4\,{\rm days}$ \citep{Abt2002}.
This corresponds to a rotational timescale of
\begin{equation}
t_{\rm rot}=\frac{1}{2 \Omega} \sim 3\times 10^4\, {\rm s},
\end{equation}
where $\Omega=2\pi/T$ is the rotation frequency.
Because $t_{\rm b}\ll t_{\rm rot} \ll t_{\rm c}$, rotation is subdominant in the radiative zone, but plays an important role in convection.
The Coriolis force leads to anisotropic convective motions.
The flow organizes into cells aligned with the rotation axis with typical vertical extents of $R_c$, but typical horizontal sizes of $Ro \, R_c$, where $Ro$ is the Rossby number.
While these anisotropic cells carry the stellar luminosity more efficiently as non-rotation convection, their relevant length scale is smaller than $R_c$; since $t_{\rm rot}\ll t_{\rm c}$ we can estimate $Ro$ using the estimates of \citet{Vasil2021},
\begin{equation}
Ro = \left(\frac{u_c}{2\Omega R_c}\right)^{3/5}\sim 0.1.
\end{equation}

This means we expect core convection to be strongly influenced by rotation, with a length scale perpendicular to the rotation axis of $0.1\, R_c$, rather than the large-scale dipolar flow which dominates in non-rotating simulations.
This demonstrates the importance of rotation on core convection.
Because the rotation time is short relative to the convection time, numerical simulations must ensure they conserve angular momentum \citep{Jones2011}.

\subsubsection{Magnetism}

The cores of intermediate- and high-mass stars are highly conductive plasmas, which can generate magnetic fields via the dynamo effect \citep{moffatt2019}.
Turbulence easily generates small-scale dynamos, tangled magnetic fields on the viscous length scale which are in equipartition with the small-scale velocity.
Rotation organizes the convective flows, and can lead to a large-scale dynamo magnetic fields (e.g., in the Sun).
These large-scale fields can sometimes be superequipartition, i.e., have larger magnetic energy than kinetic energy~\citep{Yadav2016}.
Large fields would likely impact energy transport, convective boundary mixing, and the generation of internal gravity waves by core convection.

Recently, asteroseismology has been used to infer the presence of strong magnetic fields in the cores of intermediate- and high-mass stars.
\citet{Cantiello2016} estimated that magnetic fields of strength $\sim 10^5\, {\rm G}$ would have a significant influence on $g$-mode asteroseismology of $\gamma$-Dor and slowly pulsating B (SPB) stars.
Indeed, \citep{Lecoanet2022} recently inferred the presence of a $\approx$$500\,{\rm kG}$ magnetic field near the convective core of the SPB star, HD 43317.
It is unclear if this magnetic field was dynamo-generated or is of fossil origin.
As these stars evolve off the main sequence, their cores become stably stratified, but they can maintain some of their core magnetic field generated on the main sequence.
Such magnetic fields may explain depressed dipole modes in red giant branch (RGB) stars \citep{Stello2016}, and has been used to estimate a core magnetic field of $\approx$$10^7\,{\rm G}$ in the RGB star KIC8561221~\citep{fuller2015}.
More recently, \citet{li2022} has measured 50--100 ${\rm kG}$ magnetic fields in the cores of several RGB stars.
Although magnetoasteroseismology remains in its infancy, preliminary results suggest main sequence stars with convective cores generate or maintain large-scale magnetic fields.

Magnetic fields can be included in numerical simulations of core convection by solving the magnetohydrodynamic equations.
The main difficulty in solving these equations is maintaining a divergenceless magnetic field ($\boldsymbol{\nabla}\boldsymbol{\cdot}\boldsymbol{B}=0$).
The magnetic field evolves according to the induction equation, which preserves the divergence of the magnetic field.
However, numerical errors can lead to a non-zero (exponentially growing) divergence of the magnetic field, even when initialized with a divergenceless field.
Special numerical methods are required to keep $\boldsymbol{\nabla}\boldsymbol{\cdot}\boldsymbol{B}$ small.

\subsubsection{Microphysics}

Microphysics prescriptions for nuclear reaction rates, equation of state, and opacities should be included in convection simulations.
Nuclear reactions are very slow relative to the convection time ($t_{\rm e}\gg t_{\rm c}$), so it is likely sufficient to use a radially dependent nuclear energy generation rate to drive convection.
Many simulations use an ideal gas equation of state with constant ratio of specific heats.
This is a good approximation for stars with $M\lesssim 30 M_{\odot}$, but for more massive stars, radiation pressure becomes important and should be included in the equation of state \citep{Jermyn2022a}.
The radiative luminosity produces a net cooling effect in the outer part of the convection zone, and is set by the plasma's opacity.
This opacity depends on the temperature, pressure, and chemical composition of the plasma.
Within the convection zone, the chemical composition is nearly homogeneous and the star is nearly adiabatic, so a radially dependent opacity is sufficient.
However, the temperature, pressure, and chemical dependence of the opacity are important for setting the structure of the radiative--convective boundary in thermal equilibrium \citep{Anders2023}.
Thus, studies of convective boundary mixing should include realistic opacities.

 \section{Current Approaches}
\label{sec:approaches}
Multidimensional multiphysics simulations have a long tradition in theoretical astrophysics. While the basic equations describing the physics are mostly uncontroversial, trade-offs have to be made to make their solution computationally feasible.

\subsection{Equations Describing the Problem}
A plasma in the typical density and temperature ranges of a stellar core is well described using the Navier--Stokes equations. These equations are generally applicable when the mean free path between collisions of the constituent particles is orders of magnitude smaller than the size at which the macroscopic properties, such as density, velocity, or temperature change \citep{landaulifshitz10eng}.

The Navier--Stokes equations form a system of conservation laws, which can best be seen if they are expressed in conserved variables, density~$\rho$, momentum densities $\rho u$, $\rho v$, and $\rho w$ in the three spatial dimensions, total energy density~$\rho E$, and the partial density of the $i$th species~$\rho X_i$.

The Navier--Stokes have the form (e.g., \citep{landau1987a}),
\begin{equation}
  \label{eq:navier}
\PD{\vec{U}}{t} + \PD{\vec{F}(\vec{U})}{x} + \PD{\vec{G}(\vec{U})}{y} + \PD{\vec{H}(\vec{U})}{z} = \partial_x \vec{F}^d + \partial_y \vec{G}^d + \partial_z \vec{H}^d + \vec{S}.
\end{equation}

This is expressed in terms of the flux in the Cartesian directions,
\begin{equation}
\vec{U} = \begin{pmatrix}
\rho\\
\rho u\\
\rho v\\
\rho w\\
\rho E \\
\rho X_i
\end{pmatrix},
\vec{F} = \begin{pmatrix}
\rho u\\
\rho u^2 + p\\
\rho u v\\
\rho u w\\
u(E+p)\\
\rho u X_i
\end{pmatrix},
\quad
\vec{G} = \begin{pmatrix}
\rho v\\
\rho u v\\
\rho v^2 + p\\
\rho vs. w\\
v(E+p)\\
\rho vs. X_i
\end{pmatrix},
\quad
\vec{H} = \begin{pmatrix}
\rho w\\
\rho u w\\
\rho vs. w\\
\rho w^2 + p\\
w(E+p)\\
\rho w X_i
\end{pmatrix}.
\end{equation}
The internal energy density $\rho \epsilon$ can be calculated from the total energy density using $\rho \epsilon = \rho E - \frac{1}{2} \rho \left( u^2 + v^2 + w^2 \right)$. The pressure~$p$ is a function of $\rho$, $\epsilon$, and the composition mass fraction~$X_i$ in general and is given by the equation of state. In the following, we call $\vec{F}$, $\vec{G}$, and $\vec{H}$ the \textit{hydrodynamic terms}.

The \textit{diffusive terms} take the form
\begin{align}
  \vec{F}^d &= \left(0, \tau^{xx}, \tau^{xy}, \tau^{xz}, u \tau^{xx} + vs. \tau^{xy} + w \tau^{xz} + K \partial_x T, 0\right)^T,\\
  \vec{G}^d &= \left(0, \tau^{yx}, \tau^{yy}, \tau^{yz}, u \tau^{yx} + vs. \tau^{yy} + w \tau^{yz} + K \partial_y T, 0\right)^T,\\
  \vec{H}^d &= \left(0, \tau^{zx}, \tau^{zy}, \tau^{zz}, u \tau^{zx} + vs. \tau^{zy} + w \tau^{zz} + K \partial_z T, 0\right)^T,
\end{align}
with the thermal conductivity and radiative diffusion captured by~$K$ and the components of the viscous stress tensor,
\begin{align}
  \tau^{xx} &= \frac{4}{3} \eta \partial_x u - \frac{2}{3}\eta(\partial_y vs. + \partial_z w),\\
  \tau^{yy} &= \frac{4}{3} \eta \partial_y vs. - \frac{2}{3}\eta(\partial_z w + \partial_x u),\\
  \tau^{zz} &= \frac{4}{3} \eta \partial_z w - \frac{2}{3}\eta(\partial_x u + \partial_y v),\\
  \tau^{xy} &= \tau^{yx} = \eta (\partial_y u + \partial_x v),\\
  \tau^{yz} &= \tau^{zy} = \eta (\partial_z vs. + \partial_y w),\\
  \tau^{zx} &= \tau^{xz} = \eta (\partial_x w + \partial_z u).
\end{align}
This introduces the shear viscosity~$\eta$. The bulk viscosity can be ignored in the case of a fully ionized, non-relativistic gas, which is the case relevant for core convection. To estimate the importance of viscous effects in comparison to hydrodynamics it is customary to use the Reynolds number~$\mathrm{Re} = \frac{\rho u L}{\eta}$. If this number is high, viscous effects are of minor importance and are often ignored in hydrodynamics schemes. The Navier--Stokes equations without the viscous terms are called the \textit{Euler equations}. Values in stars are typically above $10^{10}$, which means viscosity can safely be neglected. Yet, depending on the particular hydrodynamics scheme, a small amount of viscosity might be needed for the solution to converge to the physically correct one \citep{lecoanet2016a}.

As mentioned above, an equation of state (EoS) is needed to calculate the pressure from the conserved variables~$\vec{U}$. How realistic a choice of EoS is depends on the type of star. In general, the core of any star after the zero-age main sequence and before core collapse is well described by an EoS consisting of the contributions of an ideal gas of nuclei, Fermi gas electrons and positrons at arbitrary degeneracy, and a Bose gas of photons (black-body radiation). The conditions in the core allow us to assume full ionization and that all these components are in local thermodynamic equilibrium. \citet{timmes1999a} give exact expressions for all these components. While the degeneracy does not play an important role for stars on the main sequence, radiation pressure plays a major role in more massive stars even on the main sequence due to their hotter interiors. \citet{Jermyn2022a} state that the effect is negligible in stars with masses less than $9\,M_\odot$, but for stars above $20\,M_\odot$ the correction is between 30\% and 100\% depending on age.

\subsection{Numerical Solution Methods}
\subsubsection{Grid Geometry}
While the actual physics of fluid dynamics in a star is inherently three-dimensional (3D) as no symmetries can be exploited in the dynamical case, two-dimensional (2D) simulations continue to be commonly used in simulations of stellar convection. Two-dimensional domains are the simplest that support convection arising from first principles and their significantly reduced computational demand makes them appealing for test calculations and larger parameter studies. Yet, in particular, the behavior of turbulence is fundamentally different between 2D and 3D. In the 2D case, energy is transported to larger scales, which causes a flow dominated by eddies the size of the whole convective region \citep{Kraichnan1980,Boffetta2012}. In the realistic 3D case, energy is transported to smaller scale structures until a size is reached in which viscous forces dominate. This different morphology can have significant impact on transport within the convection zone, as well as convective boundary mixing.

In the 2D case, assumptions need to be made about the geometry of the third dimension. The simplest approach is to just remove the components and quantities related to the third dimension. It is effectively equivalent to the assumption that the domain extends infinitely in the third dimension. In this picture, a spherical object like a star turns into an infinite cylinder. This is typically used in the so-called annulus geometry (e.g., \citep{rogers2005a}), which simulates an equatorial slice through the star. The main downside of such an approach is that the cell areas and cell volumes do not scale with the square of radius as they should, which causes wrong results for the flux of radiation and self-gravity. Yet, if a simulation code treats these effects separately from the geometry, such as subtracting a background state from the equations and using fixed gravity calculated from the spherical case, these simulations are still very valuable as they capture the basic behavior of both convection and waves in the star.

Another approach to 2D simulations is using azimuthal symmetry. This only leaves the radial and latitudinal coordinates and means that the face areas and volumes of the cells have to be calculated accordingly. This has the big advantage that it captures the actual geometry of a spherical star in two spatial dimensions. This method is used by some of the codes which have a 2D mode, such as MUSIC or SLH. 

Three-dimensional simulations of convection in stars typically use either a Cartesian or spherical mesh. The Cartesian case has the advantage that it is very simple in terms of implementation and it is trivial to have the same resolution everywhere. Additionally, it is very amenable to be combined with mesh-refining techniques. One major downside of this approach is that it represents a box inside a star, which means that the boundaries of the simulation do not align with the natural structure of the star, which is given by radial layers. This problem can be avoided using an immersed boundary at the expense of additional complexity. Here the boundary condition is imposed on a spherical shell (or possibly another shape) within the Cartesian simulation domain. Typically, no calculations are performed outside this boundary.

If spherical coordinates are used, they are often implemented as a spherical shell. This is to avoid the singularity at the origin. While the spherical shell captures the spherical nature of a star very well and makes it easy to impose spherical boundary conditions, existence of an inner boundary possibly has an impact on the result of the simulation because it influences or even prevents flows through the center of the star.

\subsubsection{Temporal Discretization}
\label{sec:temporal-discretization}
To solve the underlying PDEs (partial differential equations) describing the stellar interior using a numerical algorithm, the continuous solution has to represented using a finite number of values. This mapping process is called \textit{discretization}. For hyperbolic PDEs, such as the Navier--Stokes equations, it is convenient to distinguish between the discretization of the spatial derivatives and the time derivatives. A common approach is the \textit{method of lines}, in which the spatial discretization is performed first and this semi-discrete system (i.e., still containing the continuous time derivative) is then evolved in time using methods for ordinary differential equations (ODEs). This makes the various methods developed for ODEs instantly available for the time-stepping of the PDEs.

There are several factors to consider in choosing the time-stepping method. A crucial one is stability. An unstable method will amplify small fluctuations until they completely dominate the solution. \citet{courant1928a} showed a generally necessary condition for the stability of the discretization of a PDE is that the numerical domain of dependence of a time step is greater or equal than the physical domain of dependence. The numerical domain of dependence refers to which parts of the domain the solution in the next step is dependent on, for example two grid cells in every direction. This is also called the stencil and only depends on the numerical method used. The physical domain of dependence on the other hand is a property of the PDEs equal to the distance information, for example a sound wave, travels during a time step.

ODE solvers for time stepping are typically grouped into explicit and implicit methods. A method is called \textit{explicit} if the new time step can be calculated using an explicit expression only depending on information from the previous time steps. The big advantage of such a method is that no additional algebra needs to be performed to solve a, possibly nonlinear, system of equations, which is computationally very expensive.
A method is called \textit{implicit} if the update is only given by an implicit, possibly nonlinear, system of equations involving information from the previous and the new time step. Depending on the equations solved this system can be linear or nonlinear. In the former case the matrix representing the system is usually of a simple structure, which is suited for direct solution methods. Examples of a linear system are the terms related to diffusion or viscosity or a linearized form of the equations of hydrodynamics. Even solving a linear system is computationally more demanding than an explicit method, partly due to the need for global communication. If a nonlinear system, such as the one that arises from the full equations of hydrodynamics, is involved, the computational cost is even higher. Here, a typical approach is to use an iterative method, which often relies on knowing the derivative of the expression. While implicit time steps are more expensive, the larger time step size can offset the total computational cost for simulation a fixed physical.

Apart from stability there is also the matter of accuracy. This is typically characterized by the method's order of accuracy. A scheme is called  order $s$ if its error reduces as $\mathcal{O}(\Delta t^s)$ as the time step $\Delta t$ is reduced to 0. The higher accuracy of high-order methods comes at the price of combining multiple evaluations of the spatial discretization for the calculation of a single new time step. This can either happen by computing several subintervals, called a \textit{single-step method}, or by combining information from previous time steps, called a \textit{multi-step method}. There are implicit and explicit variants of either of these.

Just for the hydrodynamics part of the equations the time step~$\Delta t$ scales with the minimum of $\Delta x/(u + c_s)$, which the grid spacing $\Delta x$, the fluid velocity~$u$, and the sound speed~$c_s$. The diffusive terms scale with $\Delta x^2$. This means that for explicit schemes increasing the resolution will cause the time step to be seriously restricted by the diffusive terms. That is why many codes use a hybrid implicit--explicit (IMEX) approach, which calculates the hydrodynamics terms explicitly and the diffusive terms implicitly. Additionally, a diffusion equation only results in a linear system to be solved, which is significantly cheaper than a nonlinear system. If the time step does not change, the bulk of the computational work can even reused, which is something that some codes exploit for efficiency by only changing the time step if conditions change significantly.

The size of the $\Delta t$ depends on the spatial and temporal discretization method. Explicit algorithms typically use the time step given by the CFL criterion, as this stability constraint is much more restrictive than the threshold for time-stepping accuracy that is typically chosen considering the other uncertainties in the simulation. This is different for implicit methods, which have a very large stability threshold or none at all. Here, the time step is chosen based on accuracy considerations. A typical choice is the advective CFL condition, which just considers the fluid velocity, but not the sound waves (e.g., \citep{goffrey2017a,horst2020a}). \citet{leveque2002a} describes time-stepping methods and the associated stability criteria in more detail.

The concrete choice of an ideal time-stepping method depends on the characteristics of the flow. Generally, the additional effort of implicit methods will only be beneficial if they promise significantly increased time steps. This is almost always the case for diffusion terms, but for hydrodynamics that is only true in the case of low Mach numbers.

Fully implicit 3D simulations of hydrodynamics have only recently become feasible due to advances in computing and significant development efforts \citep{miczek2015a,viallet2016a}. The break-even point at which the fully implicit approach is more efficient than a fully explicit approach is reported at Mach numbers of about $10^{-2}$ \citep{viallet2016a}. Core convection, especially in early evolutionary phases, such as core hydrogen burning, is in this regime, which is why both approaches are used in practice. Even if an explicit method is used, the diffusion terms are almost always calculated using an implicit method due to the even more restrictive stability criterion.

\subsubsection{Spatial Discretization}
There is a wide choice of methods used to discretize the spatial derivatives of the Navier--Stokes equations. The methods typically used in stellar astrophysics can be categorized into three groups:

\textit{Finite-difference methods} represent each quantity by its values at discrete points and calculate the spatial derivatives using the finite-difference approximation. A linear combination of neighboring points, called the stencil, is used to calculate the derivative at a certain order of accuracy. This is a straightforward approach that is comparably easy in its implementation. One major drawback is that the finite difference approximation works poorly close to discontinuities, which naturally occur close to shocks. At the low Mach numbers present in most situations where core convection occurs this is not an issue, but it can be a significant problem in late evolutionary stages, where the Mach numbers are closer to 1.

The class of \textit{finite-volume schemes} is designed with flow discontinuities in mind. The idea is to partition the computational domain into non-overlapping volumes, called cells or zones, and store the averages of the conserved variables on these cells. By the nature of Equation (\ref{eq:navier}) being a conservation law the changes of these averages can be determined by calculating fluxes between neighboring cells. Results from this approach fulfill the conservation of mass, momentum, energy, and chemical species down to machine precision by construction.

There are two main choices to be made for a finite-volume method. The first is the choice of numerical flux function. This is a function of the fluid state left and right of an interface that determines the flux of the conserved quantities between these neighboring cells. A trade-off has to be made between accurately representing the Navier--Stokes equations and stabilizing the flow enough to prevent being dominated by numerical noise. Many of the usual choices were designed with flow discontinuities, such as shocks, in mind and work excellently for representing these. The usually smooth flows at low Mach numbers pose a challenge to these flux functions, with the solution often completely dominated by numerical artifacts \citep{miczek2015a,barsukow2017a,horst2020a}. The reason is that the artificial viscosity terms that are added to a scheme to make it stable do not change with the same power of the Mach number as the physical terms from the PDE. There is a large variety of methods that change this scaling behavior and make sure the physical solution is also retained at low Mach numbers (e.g., \citep{liou2006a,li2008a,rieper2011a,miczek2015a,minoshima2021a}).

The other aspect concerns the spatial order of the scheme. Just assuming that the cell values are just constant over a cell and, therefore, using the respective cell averages as inputs to the numerical flux function results in a scheme that is first order accurate in space. A procedure called \textit{reconstruction} is used to increase the spatial order of the scheme. Here, the value in a cell and its neighbors is used to replace the assumption of constant values in a cell with another function, typically a linear or parabolic function. Using a linear function will make the method second order in space, a parabolic method would make it third order in space. Linear or higher order reconstruction has the problem that create oscillations in the reconstructed state, that are not there in the underlying flow. These can grow to a degree that they completely destroy the result. This is mostly a problem at Mach numbers close to 1, where it is remedied by introducing so-called limiters. These locally reduce the order of the scheme to keep it stable. While a common component of most finite-volume codes, the low Mach number flows typically present in core convection can be simulated without using limiters \citep{horst2020a}. \citet{toro2009a} and \citet{leveque2002a} give an overview of the various aspects of finite-volume methods.

A very different approach to discretization is given by \textit{spectral methods}. These rely on representing the state of the fluid as a linear combination of a set of basis functions instead of values at specific points or averages over cells. Typical basis functions are the Fourier basis, spherical harmonics, or Chebyshev polynomials. The big advantage of many spectral bases is that it often reduces the calculation of a derivative to a simple algebraic problem. In the example of the Fourier basis, a function $f(x) = \sum_{k=0}^N f_k e^{ikx}$, represented using the coefficients $f_k$, can be differentiated by simply multiplying the coefficients with $ik$. A well-chosen basis can also remove grid singularities, which limit the time step or cause artificial boundaries. For example, spherical harmonics do not have a singularity at the poles and Zernike polynomials avoid the singularity at the center of the mesh.

Solving the Navier--Stokes equations completely in spectral space results in systems of equations, which are often unfavorable for efficient computation. A common approach followed instead is a \textit{pseudo-spectral} method, in which the linear terms are solved in spectral space and the nonlinear terms are solved in physical space, with the appropriate transforms being performed as needed. \citet{glatzmaier2013a} gives a thorough introduction to using pseudo-spectral methods in the context of stellar convection. This approach was first introduced to the astrophysics community through the code of \citet{glatzmaier1984a} and has recently been significantly improved in terms of numerical efficiency in the ASH \citep{clune1999a}, MagIC \citep{wicht2002a,gastine2012a}, and Rayleigh \citep{featherstone2016a} codes. The Dedalus code \citep{Burns2020} follows an even more general approach by implementing a generic spectral solver for PDEs, which can be be used for solving the equations of (magneto-)hydrodynamics among other applications.

\subsubsection{Table of Simulation Codes}
There is a multitude of codes suited for the simulation of core convection. We list a selection of these codes in Table~\ref{tab:codes}, along with their key properties according to the criteria mentioned above. This table does not strictly list only codes which have been applied to core convection in a publication, but also codes which support all the necessary methods to address it. \citet{Kupka2017} provide an alternative list of codes, that also includes codes that have only been used in shell convection so far.

The Seven-League Hydro code (SLH) is a finite-volume code solving the fully compressible Euler equations, optionally including magnetic fields. Its main distinguishing feature is its capability for fully implicit time stepping using a variety of implicit Runge--Kutta methods and the availability of several low Mach number flux functions and well-balanced gravity \citep{edelmann2021a}. It uses a structured curvilinear mesh, which can be used to represent Cartesian, cylindrical, or spherical coordinates, but also enables more uncommon configurations, such as a cubed-sphere mesh \citep{calhoun2008a}. Additionally, it supports magnetic fields and arbitrary nuclear reaction networks. It has been used for simulating core convection in a $3\,M_\odot$ ZAMS star \citep{horst2020a}.

The MUSIC code also solves the fully compressible Euler equations using a finite-volume method with implicit time-stepping. It has a fully implicit approach using the Crank--Nicolson method. An important difference from most of the other codes discussed here is its use of a staggered grid discretization. That means that density and energy are stored on cell centers, but velocity is stored at the cell interfaces. This gives the code good properties for low Mach number flows even without resorting to special low Mach number flux functions \citep{viallet2016a,goffrey2017a}. It supports Cartesian and spherical geometry, with the option for a 2D spherical mesh with azimuthal symmetry. \citet{baraffe2023a} used MUSIC to study overshooting in convective cores.

Athena++ is a finite-volume code that solves the fully compressible Euler equations, with the option of magnetic fields and general relativity. It uses block-based adaptive mesh refinement (AMR) on a mesh using Cartesian, cylindrical, or spherical coordinates, which makes it possible to refine the mesh just where it is needed to track smaller scale physics. Being a fully explicit code would be severely limiting the time step for Athena++ if diffusive physics is included. This is avoided by introducing a Runge--Kutta--Legendre super-time-stepping scheme for the diffusive terms and computing the other terms with a normal explicit scheme in an operator split fashion.

PROMPI \textls[-15]{is a successor of the Prometheus code \citep{fryxell1991a} that has been parallelized using MPI (Message Passing Interface). It is a finite-volume code using the piecewise-parabolic method (PPM) \citep{colella1984a} with extensions for a general equation of state \citep{colella1985a}. \mbox{\citet{Meakin2007}}} used it in the simulation of hydrogen core burning in a $23M_\odot$ star and in oxygen and carbon shell burning \citep{viallet2013a,cristini2017a}.

FLASH \citep{fryxell2000a} is another code using PPM. It provides AMR and a rich set of additional physics, including nuclear burning and magnetohydrodynamics. While originally designed with stellar explosions in mind, it has been applied to shell burning in late evolutionary stages \citep{couch2015a,fields2021a}.

PPMstar solves the fully compressible Euler equations using a finite-volume scheme on a uniform Cartesian mesh with explicit time-stepping. It is highly optimized for extremely large HPC systems, allowing it to run at very high resolutions, such as $1728^3$. In order to trace the mixing of species to high accuracy it uses the Piecewise-Parabolic Boltzmann (PPB) scheme \citep{woodward2013a}, although this limits the code to tracking two fluids at most. PPMstar has been used in a study of core convection and wave excitation in a $25\,M_\odot$ main-sequence star \citep{herwig2023a,thompson2023a}.

Castro is another finite-volume hydrodynamics code using explicit time-stepping. It supports AMR through the AMRex framework \citep{zhang2019a}. It contains a rich set of additional physics, MHD, a general equation of state, a nuclear reaction network, and radiation hydrodynamics through a multigroup flux-limited diffusion. It was mostly used in the context of explosive phenomena, such as supernovae, but is also applicable for convection simulations, especially if it involves higher Mach numbers.

The ENZO code employs finite-volume (magneto-)hydrodynamics with AMR. It has been used for the study of turbulence (e.g., \citep{grete2017a}), although this was mostly focused on supersonic turbulence, which is not relevant for stellar convection. \citet{hristov2018a} used it to study the impact of magnetic fields on subsonic deflagration fronts in Type Ia supernovae.

MAESTROex uses the same AMR finite-volume mesh as Castro, but changes the equations of hydrodynamics to be efficiently solved at low Mach numbers. Its version of the pseudo-incompressible equations have been modified to work with stratified atmospheres, nuclear flames, and general equations of state (e.g., \citep{Vasil2013}). As an addition to the mesh of Castro, MAESTROex is calculating a one-dimensional base state in either plane-parallel or spherical geometry, which is needed in its version of the pseudo-incompressible equations. The time-stepping is performed using an explicit predictor--corrector method, with a special splitting approach to make sure the divergence constraints are fulfilled after applying reaction terms. The downside of the pseudo-incompressible approach is that it does not converge to the correct compressible solution as Mach numbers approach 1, though this is not an important limitation for studying core convection.

The PENCIL code is a fully compressible, finite-difference code, supporting Cartesian, cylindrical, and spherical coordinates. It uses explicit time-stepping with a third-order Runge--Kutta method and uses a sixth-order spatial discretization by default. While PENCIL has not been used in a study of core convection to date, its high-order discretization and capability for combustion, self-gravity, and MHD make it suited for the task, in principle.

\textls[-30]{The ASH code solves the Boussinesq or the anelastic equations of (magneto-)hydrodynamics} using a pseudo-spectral discretization. It uses spherical harmonics for the angular direction and one or more stacked domains of Chebyshev polynomials for the radial direction. The temporal discretization is split between an explicit Adams--Bashforth method for the nonlinear terms and an implicit Crank--Nicolson method for the implicit terms. 

The Rayleigh code uses the same type of spatial discretization and time-stepping as the ASH code, but changes the parallelization to use a two-dimensional domain decomposition instead of the one-dimensional radial decomposition used in ASH. This enables Rayleigh to scale efficiently  up to at least $10^4$ CPU cores \citep{matsui2016a}.

The SPIN code follows the same type of discretization, but uses finite differences for the radial direction. This allows the radial mesh to be adjusted to the underlying stellar model. The code was used for simulations of core convection in a 3\,$M_\odot$ ZAMS star with an artificially increased luminosity \citep{Edelmann2019}.

The MagIC code also solves the anelastic or Boussinesq equations on a spherical shell using spherical harmonics. For the radial discretization it offers the choice of using Chebyshev polynomials or finite differences. MagIC also adopts the IMEX approach for time-stepping, but, in contrast to the previously discussed pseudo-spectral codes, it offers a larger variety of methods, including single-step and multi-step methods.

The Dedalus pseudo-spectral framework can solve nearly arbitrary partial differential equations using spectral methods.
The user specifies the equations to be solved in plain text.
Dedalus can solve the Boussinesq, anelastic, generalized pseudo-incompressible, and fully compressible Navier--Stokes equations \citep{Lecoanet2014}.
It has been used to solve both hydrodynamic and magnetohydrodynamic equations, and existing simulations include other physical effects, such as self-gravity and nuclear reaction networks.
Dedalus can solve equations in Cartesian, cylindrical, or spherical geometry, typically with static mesh refinement in a single (``radial'') direction.
For spherical geometry, the angular variation of simulation variables is represented with spin-weighted spherical harmonics, and simulations can be run in full ball geometry (down to $r=0$) using generalized Zernike polynomials for radial basis functions \citep{Vasil2019,Lecoanet2019}, or in spherical shells using Jacobi polynomials for radial basis functions.
Several different multi-stage and multi-step implicit-explicit timestepping methods are implemented in Dedalus.
When the user specifies the equations they wish to solve, all terms on the left-hand side of the equals sign are treated implicitly, all terms on the right-hand side of the equals sign are treated explicitly.
Dedalus can only implicitly timestep terms which are linear in the simulation variables.
In particular, Dedalus can solve the fully compressible equations with implicit timestepping of the sound waves.

\startlandscape

\begin{table}[H]
\tablesize{\fontsize{8}{8}\selectfont} 
  \caption{\label{tab:codes}This table lists codes, which are suitable for the simulation of core convection. They are characterized by the properties introduced in Section~\ref{sec:approaches}.\\  Abbreviations: FD (finite-difference), FV (finite-volume), PS (pseudo-spectral), AMR (adaptive mesh refinement), IMEX (implicit-explicit), nuc.\ reac.\ (nuclear reactions), GR (general relativity), FL diff. (flux-limited diffusion).}
 
		\newcolumntype{C}{>{\centering\arraybackslash}X}
		\begin{tabularx}{\textwidth}{CCCCCCCC}
			\toprule
      \textbf{Name}	& \textbf{(Magneto)-Hydrodynamics} & \textbf{EoS} & \textbf{Additional Physics} & \textbf{Spatial Discretization}	& \textbf{Time Discretization} & \textbf{Select Publications} & \textbf{Licence} \\
			\midrule
      SLH		& fully compressible & general gas & multi-species, nuc. reac., self-gravity, MHD & FV (curvilinear) & explicit or implicit & \cite{miczek2015a,horst2020a} & proprietary \\
      MUSIC		& fully compressible & general gas  & multi-species & FV (Cartesian or spherical) & implicit & 
      \cite{goffrey2017a,baraffe2023a} & proprietary \\
      Athena++ & fully compressible & general gas &  multi-species, self-gravity, MHD, GR & FV (Cartesian, cylindrical or spherical), AMR & explicit, super-time-stepping & \cite{stone2020a} & free (BSD 3-clause)\\
      PROMPI & fully compressible & general gas & multi-species, nuc. reac. & FV (Cartesian, cylindrical or spherical) & explicit & \cite{Meakin2007} & proprietary\\
      FLASH & fully compressible & general gas & multi-species, nuc. reac., self-gravity, MHD & FV (Cartesian, cylindrical or spherical), AMR & explicit & \cite{fryxell2000a} & available on request\\
      PPMstar & fully compressible & ideal gas & two-fluid & FV (Cartesian) & explicit & \cite{woodward2013a,Woodward2019,herwig2023a,thompson2023a} & proprietary \\
      CASTRO & fully compressible & general gas & multi-species, MHD, self-gravity, nuc. reac, FL diff. & FV (Cartesian), AMR & explicit & \cite{almgren2020a} & free (BSD 3-clause)\\
      MAESTROex & generalized pseudo-incompressible & general gas & multi-species, self-gravity, nuc. reac. & FV (Cartesian), AMR & explicit & \cite{fan2019a,fan2019b} & free (BSD 3-clause)\\
      ENZO & fully compressible & general gas & MHD, self-gravity, nuc. reac., radiation & FV (Cartesian), AMR & explicit & \cite{bryan2014a,brummel-smith2019a} & free (BSD 3-clause)\\
      PENCIL & fully compressible & partially ionized, ideal gas & MHD, self-gravity & FD & RK & \cite{pencil2021} & free (GPLv2) \\
      ASH & anelastic or Boussinesq & ideal gas & MHD & PS (spherical shell) & IMEX & \cite{clune1999a,Browning2004,Augustson2016} & proprietary \\
      Rayleigh & anelastic or Boussinesq & ideal gas & multi-species, MHD & PS (spherical shell) & IMEX & \cite{featherstone2016a,matsui2016a,featherstone2022a} & free (GPLv3)\\
      SPIN & anelastic & ideal gas & MHD & PS (spherical shell), FD & IMEX & \cite{Edelmann2019} & proprietary\\
      MagIC & anelastic or Boussinesq & ideal gas & multi-species, MHD & PS (spherical shell), PS or FD (radial) & IMEX & \cite{wicht2002a,gastine2012a} & free (GPLv3)\\
      Dedalus & fully compressible, pseudo-incompressible, anelastic, or Boussinesq & ideal gas & multi-species, MHD & PS (spherical, full ball or shell), cylindrical, Cartesian, SMR & IMEX & \cite{Burns2020} & free (GPLv3) \\
			\bottomrule
		\end{tabularx}
 
\end{table}
\finishlandscape

 \section{State-of-the-Art Results}

\subsection{Convective Boundary Mixing}

\begin{figure}[H]
\begin{adjustwidth}{-\extralength}{0cm}
 \centering 
\includegraphics[width=1.1\textwidth]{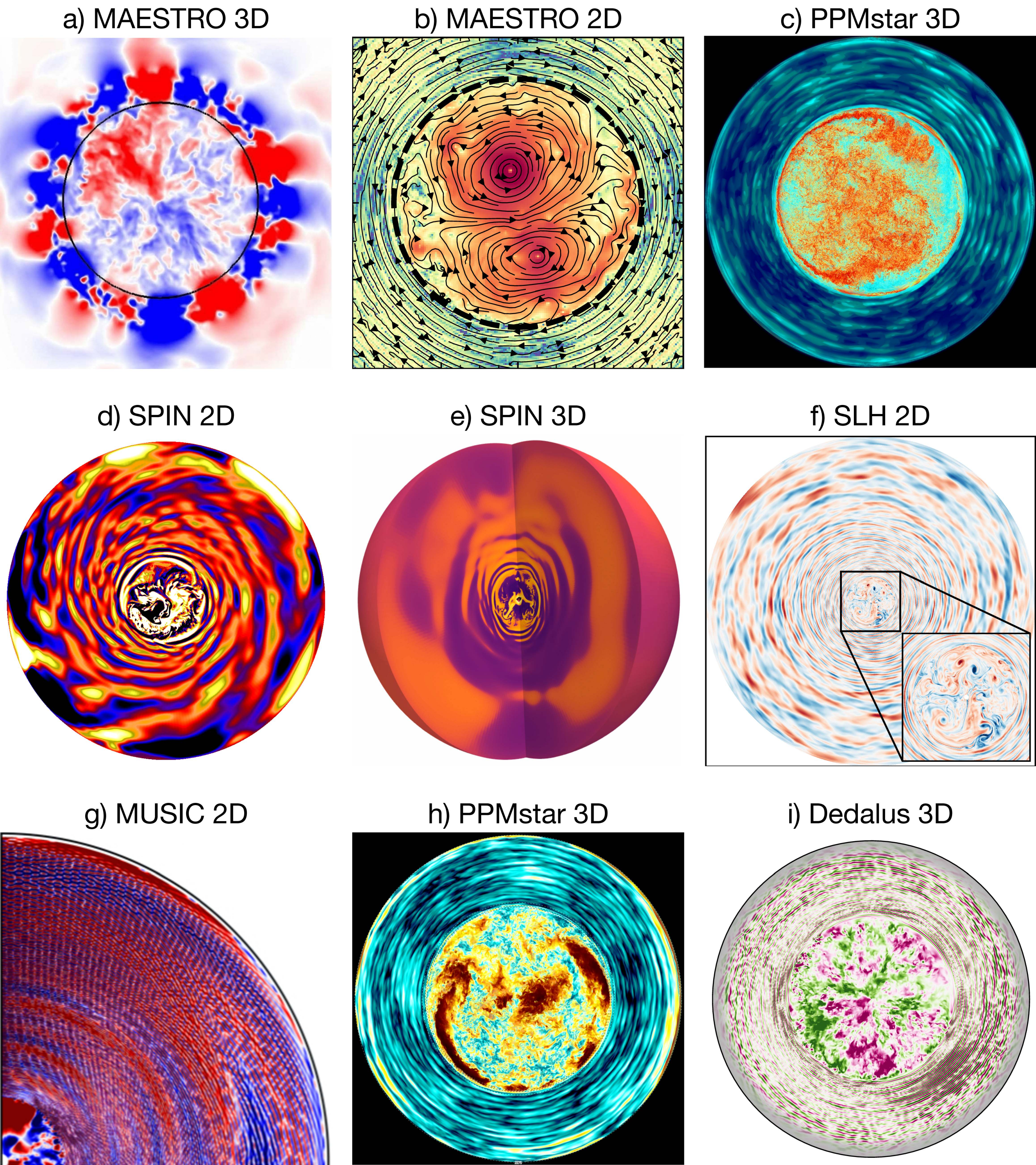}
\end{adjustwidth}
\caption{\label{fig:snapshots}\newtext{Snapshots from simulations of core convection. Top row includes simulations used to study convective boundary mixing; bottom two rows includes simulations used to study internal gravity waves. (\textbf{a}) Radial velocity from \citet{Gilet2013}; \copyright\, AAS. Reproduced with permission. (\textbf{b}) Velocity magnitude from \mbox{\citet{Higl2021}}. (\textbf{c}) Vorticity magnitude from \citet{Mao2023}. (\textbf{d}) Vorticity from \citet{Rogers2013}; \copyright\, AAS. Reproduced with permission. (\textbf{e}) Temperature fluctuations from \citet{Edelmann2019}; \copyright\, AAS. Reproduced with permission. (\textbf{f}) Temperature fluctuations from \citet{horst2020a}; \copyright\, ESO. Reproduced with permission. (\textbf{g}) Radial velocity from \citet{LeSaux2023}; we only reproduce the northern hemisphere of their axisymmetric simulation. (\textbf{h}) Horizontal velocity magnitude from \citet{herwig2023a}. (\textbf{i}) Radial velocity from \citet{Anders2023b}.}}
\end{figure}

Mixing at the radiative--convective boundary can mix fresh fuel into convective cores, increasing main-sequence lifetimes.
This mixing encompasses several different physical mechanisms including (i) growth of the convection zone (``entrainment''); (ii) transport of fluid across the radiative--convective boundary (``overshoot''); and (iii) developing an adiabatic region outside the Schwarzschild boundary (``penetration'').
These mechanisms are discussed in depth in \citet{Anders2023}, along with observational, theoretical, and numerical evidence for each mechanism.
Here, we briefly summarize the numerical studies of convective boundary mixing in simulations of core convection. Figure~\ref{fig:snapshots} shows snapshots of core convection simulations produced with various codes.

Even the earliest numerical simulations of core convection showed convective boundary mixing.
\citet{Deupree2000} ran low-resolution 2D simulations of stars with masses ranging from $1.2M_\odot$ to $20M_\odot$, and measured overshoot lengths of $0.3-0.5H_P$ by evolving passive tracer fields.
Several years later, \citet{Browning2004} ran low-resolution 3D ASH simulations of a $2M_\odot$ A star, finding evidence of both convective penetration and overshoot, with a typical overshoot length of $0.2H_P$.
In both cases, the authors note that increasing the resolution decreases the overshoot length; furthermore, \citet{Browning2004} decreased the stable stratification in the radiative zone to decrease numerical costs.
These early simulations show much a larger overshoot than both higher resolution simulations, and theoretical predictions for realistic stars, suggesting viscosity/numerical resolution played a significant role in convective boundary mixing.

Other early simulations with PROMPI measured the entrainment rate in a $23M_\odot$ model with luminosity boosted by 10 using 2D and 3D wedges containing only the outer third of the convection zone \citep{Meakin2007}.
However, the entrainment rate in these simulations is orders of magnitude higher than more recent, higher resolution simulations.
For instance, \mbox{\citet{Gilet2013}} ran simulations of a $15M_\odot$ star using Cartesian geometry with the MAESTRO code.
In their simulations, they calculate the bulk Richardson number $Ri_b= 3\times 10^5$, much larger than the $Ri_b= 60$ of \citet{Meakin2007}, but in line with the theoretical expectation $Ri_b\sim (t_{\rm c}/t_{\rm b})^2\sim 10^6$.
\citet{Gilet2013} calculate the mass entrainment rate by measuring the change in mass of the convective core.
Using high-resolution PPMstar simulations of a $25M_\odot$ star with luminosity boosted by different factors, \mbox{\citet{Woodward2019}} found the mass entrainment rate scales with $u_c\sim L^{1/3}$, consistent with \mbox{\citet{Gilet2013}} when extrapolated down to the realistic luminosity.
On the other hand, \mbox{\citet{Higl2021}} ran a series of 2D simulations of 1.3--3.5$M_\odot$ stars using Cartesian grids, and found mass entrainment rates orders of magnitude larger than \citet{Gilet2013} and \mbox{\citet{Woodward2019}}, when scaled appropriately for luminosity.
Although convective velocities appear to be somewhat larger in 2D simulations than 3D simulations, there is still a substantial discrepancy between entrainment rates calculated by different groups, even when using the same code (e.g., MAESTRO).

In addition to entrainment rates, \citet{Higl2021}, they also try to parameterize the convective boundary mixing in terms of an exponential overshoot parameter $f_{ov}$.
For this, they run a series of 2D hydrodynamic simulations initialized from 1D stellar structure models with different $f_{ov}$.
If the $f_{ov}$ from the stellar structure model is too large, they find there is negligible mass entrainment in their 2D hydrodynamic simulations.
This indicates the hydrodynamic simulation is not able to support such an extended convection zone, which allows them to place upper bounds on $f_{ov}$ which are $\mathcal{O}(10^{-2})$.
This is consistent with the penetrative convection mechanism, and similar to the simulations of \mbox{\citet{Anders2022penetration}}; and, indeed, \citet{Higl2021} see direct evidence of convective penetration in their simulations.

Other efforts to parameterize convective boundary mixing include a series of papers analyzing simulations using the anelastic approximation of \mbox{\citet{rogers2005a}}.
First, \citet{Rogers2013} describes a series of rotating 2D simulations in cylindrical geometry of a $3M_\odot$ zero-age main sequence star with artificially boosted luminosity.
They calculate an overshoot distance based on the first sign change in the average kinetic energy flux outside the convection zone, with typically lengths of $0.1-0.5H_P$.
They also note their simulations have significantly superadiabatic stratification outside the convection zone---suggesting convective penetration---again extending to lengths up to $\sim$$0.5H_P$.
\mbox{\citet{Edelmann2019}} ran similar simulations of a $3M_\odot$ zero-age main sequence star with artificially boosted luminosity, but in 3D spherical geometry.
They measure convective penetration by measuring the instantaneous distance between the surface of zero radial velocity and the convection zone boundary, which was most likely to $\sim$$0.5H_P$; however, they do not report if this region is adiabatic.
Recently, \citet{Varghese2023} derived chemical diffusivites using passive tracer particles in a series of 2D simulations in polar coordinates of stars with masses 3--20$M_\odot$.
Outside the convection zone, they found the diffusivity dropped as a Gaussian with length scale 0.1--0.2 times the pressure scale height.
This is a significantly larger overshoot than \citet{Higl2021}, who measured diffusivities using passive tracer particles in the same way.
Note that the convective velocities in \citet{Varghese2023} are similar to those predicted by MLT, whereas \citet{Higl2021} finds velocities $\sim$$10$ times larger; one would expect for the simulations with large convective velocities to have large overshoot lengths.
There remain order-of-magnitude discrepancies in the convective overshoot length when using similar analyzes on 2D simulations solving similar equations (with filtered sound waves), with similar background stellar models.

Very recently, \citet{baraffe2023a} ran a series of 2D meridional simulations (e.g., \mbox{$\partial/\partial\phi=0$}) of the fully compressible Euler equations with MUSIC.
They simulate stars near the zero-age main sequence with masses ranging from $3M_\odot$ to $20M_\odot$.
While most simulations use the stellar luminosity, they also study the effect of boosting the luminosity in their $3M_\odot$ simulations.
They measure a convective overshooting length scale by calculating the distance from the radiative--convective boundary to the first zero of the radial heat flux, $\ell_{\delta T}$.
Following the extreme value statistics arguments of \citet{Pratt2017} and \mbox{\citet{Baraffe2021}}, they equate the overshooting length to the time average of the angular maximum of $\ell_{\delta T}$.
They find that the overshooting length is proportional to the convective velocity and to the square root of the radius of the convection zone.
For zero-age main sequence stars from $3M_\odot$ to $20M_\odot$, this corresponds to $\sim$$0.05$ to $\sim$$0.2$ pressure scale heights, similar to overshoot lengths measured with other methods described above.
They also find nearly adiabatic regions outside the convection, as predicted by theories of convective penetration, but did not run the simulations long enough to determine the equilibrium size of the penetration~zone.

Even more recently, \citet{herwig2023a} studied convection in a $25M_\odot$ star near the main sequence.
Their simulations with PPMstar include $r\lesssim 0.5 R_\star$ and use Cartesian geometry.
Because their code explicitly resolves sound waves, they boost the stellar luminosity by different factors $\gtrsim$$10^{1.5}$; typical simulations boost the luminosity by $10^3$.
They find an entrainment rate proportional to the luminosity boosting; extrapolated to realistic luminosities they find $\dot{M}\sim10^{-11}\, {\rm cm}^2/{\rm s}$, similar what \citet{Gilet2013} found for a $15M_\odot$ star.
Beyond the convective boundary, they find diffusive mixing which they interpret as being due to shear instabilities of internal gravity waves (see below for further discussion).
\newtext{\citet{Mao2023} analyze similar PPMstar simulations which include radiative diffusion.
They find that the entrainment of \citet{herwig2023a} slows down with time, and builds up an adiabatically stratified region outside of the convection zone.
These are hallmarks of convective penetration \citep{Anders2023}.
They find that the entropy profiles evolve in the same way in simulations with a range of luminosity boosting factors, but that the evolution is faster in simulations with higher luminosities.
Although their penetration zone does not completely equilibrate in size by the end of their simulation, they use the framework of \mbox{\citet{Roxburgh1989}} to develop a method for including the effects of convective penetration in 1D stellar evolution codes.}

\subsection{Waves and Convection}

While intermediate- and high-mass stars have convective cores, their envelopes are primarily radiative (outside of a narrow surface convection zone, see Jiang and Cantiello 2023).
This radiative zone supports a range of waves.
The waves are classified by their main restoring force, $p$-modes, or sound waves, are associated with pressure forces; $g$-modes, or gravity waves, are associated with buoyancy forces; and $r$-modes, or Rossby waves, are associated with Coriolis forces.
All three types of waves are excited by convection.
If the waves have long lifetimes and reach sufficient amplitude, they may be detectable at the surface via asteroseismology (e.g., \citep{Bowman2019}).
The properties of such waves could provide insights into the properties of core convection.
Furthermore, waves transport chemicals and angular momentum, affects which should be included in stellar evolution models.
In this section, we will first briefly review some theoretical aspects of wave excitation and propagation, then discuss results from multidimensional simulations of core convection, as well as their implications for chemical and angular momentum transport.

\subsubsection{Theoretical Considerations}

Waves generated by convection are interesting because they influence parts of a star far from the convection zone.
If a wave is generated in the convection zone and damps deep within the radiative zone, it transports angular momentum from the convection zone to the damping location.
Waves which can propagate far from the convection zone without significant damping have the largest influence on stellar structure and are the best candidates for direct detection via asteroseismology.

The impact and observability of waves excited by convection depends on the amplitude of the waves.
The wave amplitude is set by a balance between the convective excitation rate and the radiative damping rate.
Here, we will describe theoretical models of the convective excitation rate.
These two processes are analyzed separately because the convective excitation occurs in (or near) the convection zone, and is thought to be a universal property of convection, whereas the radiative damping occurs in the radiative zone.
Care must be taken when comparing predictions of excitation rates to simulations, which typically measure wave amplitudes \citep{Lecoanet2021}.

While initial work on wave generation by convection in stars made a variety of heuristic arguments \citep{Press1981,GraciaLopez1991}, the earlier Lighthill theory of wave excitation by turbulence provides a more systematic method for calculating the convective excitation rate \citep{Lighthill1952}.
In the Lighthill theory, the Navier--Stokes equations are written in the form
\begin{equation}
L.\vec{U} = N(\vec{U}),
\end{equation}
where $\vec{U}$ is a vector of perturbation variables, $L$ is a linear operator, and $N$ is a non-linear operator.
Waves follow the dispersion relation $L.\vec{U}=0$.
The idea of the Lighthill theory is that the nonlinear terms $N(\vec{U})$ arising from turbulence act as a source term for waves.
The waves are then given by the equation
\begin{equation}\label{eqn:green}
\vec{U} = \int G(\vec{x}',t';\vec{x},t) \ N(\vec{U})(\vec{x},t) \, d^3\vec{x}'dt',
\end{equation}
where $G$ is the Green's function of the linear operator $L$.
Equation~(\ref{eqn:green}) is an exact form of the Navier--Stokes equations without any approximations.

To use Equation~(\ref{eqn:green}), one typically assumes the source term for the waves $N(\vec{U})$ is dominated by convective turbulence.
This is valid if the waves are linear and do not affect the convection.
The challenge then becomes how to model the statistical properties of $N(\vec{U})$.
This approach was already used in \citet{Goldreich1977} to calculate $p$-mode excitation in the sun.
Later, \citet{Goldreich1990} used the same approach to estimate $g$-mode excitation.
A main result is that the convective excitation rate of sound waves is weaker than the convective excitation rate of gravity waves by $Ma^{13/2},$ where $Ma$ is the Mach number of the convection.
Using a typical Mach number of $10^{-3}$ (Section~\ref{sec:challenges}), we estimate excitation rate of gravity waves is higher than the excitation rate of sound waves by a factor of $\sim$$10^{20}$.
Thus, research on wave excitation by convection has focused on gravity waves rather than sound waves.

For a spherically symmetric star and neglecting rotation, the angular structure of waves are given by spherical harmonics $Y_{\ell,m}(\theta,\phi)$.
The dispersion relation $L.\vec{U}=0$ determines the radial structure of the wave as a function of $\ell$ and its angular frequency $\omega$; it is independent of $m$.
Absent damping and nonlinear effects, the wave luminosity
\begin{equation}\label{eqn:wave luminosity}
L = 4\pi r^2\rho_0 u_r p'
\end{equation}
is constant with radius, where $\rho_0(r)$ is the stellar density profile, and $u_r$, $p'$ are the radial velocity and pressure perturbation of the wave.
However, some gravity waves experience significant radiative damping.
An outward-propagating gravity wave's luminosity decays~as
\begin{equation}\label{eqn:wave damping}
L(r) = L(r_0)\exp\left(-\int_{r_0}^{r} k_\mathrm{rad}(r')\, \frac{N(r')^3 \left[\ell(\ell+1)\right]^{3/2}}{r'^3\omega^4}\,dr'\right),
\end{equation}
where $k_\mathrm{rad}$ is the radiative diffusivity.
Radiative damping is most significant for low $\omega$ and high $\ell$.
In many theories of convective excitation of waves, the mostly efficiently excited waves have low frequencies, $\omega$ close to $1/t_{\rm c}$.
For the $10M_\odot$ zero-age main sequence star considered in Section \ref{sec:challenges}, these waves would have $\omega\sim 1/t_{\rm c}\sim 10^{-6}\, {\rm Hz}$, $N\sim 1/t_{\rm b}\sim 10^{-3}\, {\rm Hz}$, $r\sim R_c\sim 6\times10^{10}\, {\rm cm}$, and $k_\mathrm{rad} \sim 3\times 10^{9} \, {\rm cm}^2/{\rm s}$.
Such an $\ell=1$ wave has a damping length of $6\times 10^7\, {\rm cm}\sim R_c/1000$.
Although these waves may be efficiently excited by convection, they play little role in stellar evolution because they damp before they can propagate far from the convection zone.
However, waves with higher frequencies $\omega=10^{-5}\, {\rm Hz}$ damp over $10 R_c$, so may be astrophysically relevant.

This illustrates the importance of the frequency (and $\ell$) dependence of the convective excitation rate.
Different assumptions about the convective source term $N(\vec{U})$ in \mbox{Equation~(\ref{eqn:green})} lead to different predictions.
\citet{Goldreich1990} assume convection can be decomposed into eddies, each of which has a strength predicted by the Kolomogorov theory of turbulence and is coherent for its local turnover time.
These assumptions predict the energy flux from convection into waves is a power-law in frequency and $\ell$,
\begin{equation}\label{eqn:GK90}
F_w = F_c\, Ma\, \left(\omega t_{\rm c}\right)^{a} \left[\ell(\ell+1\right)]^{b/2},
\end{equation}
where $F_c$ is the convective flux, and very steep power-law indices $a\approx-6.5$ and \mbox{$b\approx 3$}.
\citet{Belkacem2009} modeled the convective source term using a three-dimensional ASH simulation of solar convection.
The time-correlation function of the velocities in the simulation were well-fit by a Lorentzian, which would give a very different frequency dependence than Equation~(\ref{eqn:GK90}), which implicitly assumes a Gaussian time-correlation function.
More recently, \citet{Pincon2016} calculated the wave excitation from overshooting plumes.
They predict the energy flux from plums into waves is a Gaussian in frequency and $\ell$,
\begin{equation}\label{eqn:P16}
F_w \sim \exp\left(-\omega^2/\nu_p^2\right)\ \exp\left(-\ell(\ell+1) b^2/r^2\right),
\end{equation}
where $\nu_p^{-1}$ is the plume lifetime $\pi b^2$ is the horizontal area of the plume.
These different models give very different predictions for the frequency and $\ell$ dependence of the convective excitation rate, and, hence, wave amplitudes.
Simulations are needed to distinguish between these theories and determine whether or not they are effective descriptions of wave excitation by convection.

Up to now, we have not discussed the effects of rotation.
Rotation affects both waves (leading to inertial waves, Rossby waves, and rotationally modified gravity waves), and convection.
The wave properties are very different depending on their frequency relative to $2\Omega$, where $\Omega$ is the rotation frequency \citep{Mathis2014}.
While intermediate frequency waves with $2\Omega \ll \omega \ll N$ are only weakly perturbed by rotation, if $\omega\lesssim 2\Omega$, waves will be strongly influenced by rotation.
In particular, there are now modes which propagate as inertial waves in the convective core, but gain gravity-wave character in the radiative zone; these modes may be more easily excited by convection than non-rotating gravity waves which are evanescent in the convection zone.
This effect was included in \citet{Augustson2020}, which also uses the \citet{Stevenson1979} model for rotating convection.
They found rotation could potentially enhance the convective excitation rate by orders of magnitude.
These works show rotation likely has a major effect on both wave excitation and propagation in stars, which should be further studied both theoretically and numerically.

\subsubsection{Simulations of Wave Excitation by Core Convection}

Even early simulations of core convection \citep{Browning2004, Meakin2007} found that convective excites gravity waves, although the waves were not studied quantitatively.
Since then, several papers have studied gravity wave excitation by core convection in great detail.
\newtext{We summarize key features of these studies in Table~\ref{tab:IGW}.
As a whole, these works confirm that as the waves propagate outward in the star away from the convection zone, their amplitude seems to evolve roughly as expected by Equations~(\ref{eqn:wave luminosity}) and (\ref{eqn:wave damping}).
At high frequencies, convection excites standing modes at the $g$-mode frequencies of the star.
The early simulations of \citet{Rogers2013} found some evidence for nonlinear wave interactions, but more recent investigations (\citep{Edelmann2019,horst2020a,thompson2023a,LeSaux2023,Anders2023b}) do not see significant wave nonlinearity away from the radiative--convective boundary.}

\begin{table}[H]
  \caption{\label{tab:IGW}\newtext{Key features of studies of convective excitation of internal gravity waves. The CZ resolution is an estimate of the number of grid points across the convective zone radius.\\ Abbreviations: AN (anelastic), FC (fully compressible).}}
	\begin{adjustwidth}{-\extralength}{0cm}
		\newcolumntype{C}{>{\centering\arraybackslash}X}
		\begin{tabularx}{\fulllength}{cCCCCCC}
			\toprule
      \textbf{Paper}	& \textbf{Code} & \textbf{Equations} & \textbf{Dimensionality} & \textbf{Mass} & \textbf{Luminosity} & \textbf{CZ Resolution} \\
			\midrule
      \citet{Rogers2013} & SPIN & AN & 2D & $3M_\odot$ & $5\times 10^4\, L_\star$ & $\approx$$400$ \\
      \citet{Edelmann2019} & SPIN & AN & 3D & $3M_\odot$ & $10^6\,L_\star$ & $\approx$$400$ \\
      \citet{horst2020a} & SLH & FC & 2D & $3M_\odot$ & $10^3\, L_\star$ & $\approx$$160$ \\
      \citet{thompson2023a} & PPMstar & FC & 3D & $25M_\odot$ & $10^3\,L_\star$ & $\approx$$432$ \\
      \citet{LeSaux2023} & MUSIC & FC & 2D & $5M_\odot$ & $1 - 10^4 \ L_\star$ & $\approx$$256$ \\
      \citet{Ratnasingam2023} & SPIN & AN & 2D & $3-13 \ M_\odot$ & $\sim 10^3\, L_\star$ & $\approx$$512$ \\
      \citet{Anders2023b} & Dedalus & FC & 3D & $3-40 \ M_\odot$ & $L_\star$ & $\approx$$512$ \\
			\bottomrule
		\end{tabularx}
	\end{adjustwidth}
\end{table}

\newtext{Of these papers, only \citet{Anders2023b} studies the wave excitation rate.
They find the wave luminosity follows the power-law form of Equation~(\ref{eqn:GK90}), with $a=-6.5$ and $b=4$, very close to the predictions of \citet{Goldreich1990}.
The discrepancy in $b$ is because \citet{Goldreich1990} calculated wave excitation by convective envelopes; \citet{Lecoanet2013} showed the same theory applied to core convective predicts $b=4$.}

\newtext{The remaining papers report} quantities related to the wave amplitude.
This complicates the interpretation of these simulations \newtext{many of which} have boosted luminosities, as the wave amplitude is set by a balance between excitation and damping rates, but the excitation rate and damping rate scale differently with luminosity.
\citet{Rogers2013} calculates the kinetic energy of the waves as a function of frequency and wavenumber, $E(\omega,\ell)$.
They find it is a good approximation to write $E(\omega,\ell)=f(\omega)g(\ell)$.
Although this separable form may seem consistent with the theoretical predictions of \citep{Goldreich1990}, Equation~(\ref{eqn:GK90}) is only valid for a range of frequencies larger than $\sim\ell^{2/3}$, which makes it not separable.
They find $f(\omega)$ and $g(\ell)$ are well-fit by broken power-laws.
While the power-law exponents varied somewhat for different simulation parameters, typical dependencies were $\omega^{-1}$ at low frequencies, $\omega^{-4}$ at high frequencies, and $\ell^{-1}$ at low wavenumbers, $\ell^{-4}$ at high wavenumbers.
They also ran simulations with a range of rotation rates, but did not find that rotation significantly affected the wave spectra or amplitudes (this appears to be rather different from the theoretical results of \citet{Augustson2020}).

\citet{Edelmann2019} also analyzes the kinetic energy of the waves right outside of the convection zone as a function of frequency and wavenumber, but does not assume separability.
They instead analyze the frequency-dependence of the kinetic energy for fixed wavenumbers $\ell$.
As in \citet{Rogers2013}, they find broken power-laws, with typical dependence $\omega^{-1/2}$ at low frequencies and $\omega^{-5}$ at high frequencies, although the exponents vary somewhat with $\ell$.
They find the frequency of the break in the power laws scales linearly with $\ell$.
This frequency dependence can be represented as a sum of Gaussians of different widths and amplitudes, similar to the excitation rate predictions by overshooting plumes in \citet{Pincon2016}.

\newtext{\citet{Edelmann2019}, \citet{horst2020a}, and \citet{Ratnasingam2023} all} report the wave velocity and temperature further out in the radiative zone as a function of frequency only (e.g., summed over $\ell$).
\citet{Edelmann2019} finds $v_r\sim \omega^{-0.8}$, \citet{horst2020a} finds $v_r\sim \omega^{-0.2}$, \newtext{ and \citet{Ratnasingam2023} finds $v_r\sim\omega^{-0.25}-\omega^{-1}$, depending on the mass of the star}.
For the temperature spectrum, \citet{Edelmann2019} finds a maximum at $\sim 6\, {\rm d}^{-1}$ followed by a roughly exponential decay.
\newtext{\citet{Ratnasingam2023} finds a plateau at low frequencies, followed by a power-law decay at high frequency with $T\sim\omega^{-1.1}-\omega^{-1.65}$, depending on the mass of the star.}
To better compare to observations, \citet{horst2020a} calculates the frequency spectrum of the temperature fluctuations averaged over half the surface of their domain.
This mimics the fact that only half a star's surface is visible at a time.
They find the temperature fluctuations peak at $\sim 1\, {\rm d}^{-1}$, then decay as a $\omega^{-2}$ power-law at higher frequencies.

\citet{herwig2023a} primarily analyzes the horizontal and radial velocity spectra as a function of spherical harmonic degree $\ell$ (e.g., summed over frequencies).
They find $v_r\sim \ell^{-5/3}$ and $v_h\sim \ell^{7/16}$ at moderate $\ell$, and both decrease as $\ell^{-9/2}$ for high $\ell$.
The transition between these regimes can occur at different $\ell$, perhaps as a function of $N^2$.
\citet{thompson2023a} then analyzes the luminosity fluctuations of these waves (proportional to the temperature fluctuations).
They find the luminosity fluctuations averaged over a hemisphere (similar to a disk-integrated photometric measurement) peaks at the convective frequency of their simulation ($\sim$$2\,\upmu{\rm Hz}$), and decreases as a power law $L\sim \omega^{-2.33}$.
This is similar to the temperature fluctuation spectra of \citet{horst2020a}.

\newtext{\citet{LeSaux2023} run a series of simulations with different convective luminosities.
They calculate the wave energy flux (similar to the wave luminosity) in their simulations.
For simulations with realistic luminosities, the wave flux has a broad peak at low frequencies, and then decreases as a power law at higher frequencies.
They interpret the low-frequency excitation to be by convective plumes, and the high-frequency excitation to be by Reynolds stresses (similar to \citep{Rogers2013}).
The high-frequency behavior matches \mbox{Equation~(\ref{eqn:GK90})} with $a\approx-6.5$ (similar to \citep{Anders2023b}).
Similar work for solar-type stars also found $a\approx -6.5$ and $b\approx 4$ \citep{LeSaux2022}.
As the luminosity increases, they find the peak of the spectrum shifts to higher frequencies, and the amplitude of the spectrum increases.
Together with~\citep{LeSaux2022}, this illustrates the challenges of interpreting simulations with boosted luminosities, as radiative damping of waves is strongly frequency dependent.}

One goal of these simulations is to determine the luminosity fluctuations at the stellar surface from waves excited by convection.
\citet{Bowman2019} detected ubiquitous low-frequency variability in stars with convective cores.
The amplitude of the luminosity fluctuations is $\sim$10--$10^{3}\, \upmu{\rm mag}$ at frequencies below a characteristic frequency $\sim$$1\, {\rm d}^{-1}$, and decays as a power-law with exponent around $-2$ above the characteristic frequency.
It is difficult to compare simulations to these observations because the simulations do not extend to the surface and because it is not straightforward to determine which simulation variable corresponds to brightness variations.

In \citet{Rogers2013} and \citet{Edelmann2019}, the tangential velocity at the top boundary is used as a proxy for surface brightness variations.
There is a good match in the frequency spectra in the 2D and 3D simulations, and both are similar to the brightness variations of observed stars (see also \citep{Aerts2015}).
However, the simulations have too little power at low frequencies $\lesssim$$2\, {\rm d}^{-1}$, and the decay in power at high frequencies is more rapid in the simulations than in the observations.
The $\sim$$\omega^{-2}$ decay in the temperature spectrum of \mbox{\citet{horst2020a}} and the luminosity spectrum of \citet{thompson2023a} is similar to observations.
\mbox{\citet{thompson2023a}} notes that using the iterative pre-whitening of \mbox{\citet{Bowman2019}} removes the highest-amplitude luminosity perturbations and reproduces the Lorentzian spectra of \citet{Bowman2019}.
\newtext{\citet{Ratnasingam2023} calculated temperature fluctuations averaged over half their simulation, and including limb-darkening effects.
They found these temperature fluctuation spectra were in very good agreement with rescaled observed photometric variability spectra.}

None of the \newtext{above} papers quantitatively compare the amplitudes of the perturbations in the simulations to the amplitudes of observed variability.
The simulated amplitudes are likely unrealistically high as in each case the simulations use boosted luminosities~\citep{LeSaux2023}.
Further theoretical developments are required to ``un-boost'' \newtext{these} simulations to make more quantitative comparisons to observations.
\newtext{However, the recent work of \mbox{\citet{Anders2023b}} predict the amplitude of observed variability from simulations using realistic luminosities.
They balance the convective excitation rate from numerical simulations with the radiative damping rate from the GYRE non-adiabatic pulsation code.
They find the amplitude of internal gravity waves excited by core convection is $\lesssim$$0.1\,\upmu {\rm mag}$ for ZAMS stars with masses 3--40$M_\odot$, much lower than observed variability.}

\subsubsection{Chemical Mixing by Waves}

A major open question in stellar astrophysics is: what determines the transport of chemicals in the radiative zones of stars?
Recent asteroseismic studies have constrained chemical diffusivities in the radiative zones of stars with convective cores \citep{Pedersen2021}.
One possible source of chemical transport is waves.
The theory for chemical transport by waves is described in \citet{Jermyn2022b}.
Linear waves do not produce net transport of chemicals, but the nonlinear Stokes drift does lead to net chemical transport.
The Stokes drift is in the direction of wave propagation and has magnitude $u_s\sim u_r^2/u_g$, where $u_r$ is the radial velocity of the wave and $u_g$ is its radial group velocity.
While this drift causes transport, it is only in the direction of the wave propagation, so cannot be thought of as a diffusive process.
However, chemical species can still diffuse due to the random superposition of Stokes drifts from many waves, some of which transport outward, some of which transport inward.
Such a diffusion would be proportional to $u_s^2 \sim u_r^4$.
\citet{Jermyn2022b} presents a detailed calculation of this wave diffusivity, which is proportional to the wave velocity to the fourth power.

This theoretical model is not supported by numerical simulations.
\citet{Rogers2017} introduced passive tracer particles into 2D cylindrical anelastic simulations of a $3M_\odot$ zero-age main sequence star, similar to those of \citet{Rogers2013}, but without luminosity boosting.
They find the particles diffuse, with a diffusivity which varies with radius.
They find the diffusivity is given by $D_w=u_w^2/\tau$, where $u_w$ is the wave velocity and $\tau=1\, {\rm s}$.
The diffusivity ranges from $10^8\,{\rm cm}^2/{\rm s}$ near the convection zone to $10^{10}\, {\rm cm}^2/{\rm s}$ near the top of the simulation ($0.7R_\star$).
In a recent follow-up paper, \citet{Varghese2023} performed a similar analysis for 2D simulations with stellar masses between $3M_\odot$ and $20M_\odot$, from the zero-age main sequence to the terminal-age main sequence.
They demonstrated the chemical diffusivities are converged with respect to the number of particles and numerical choices associated with the particle particle tracking.
They find that across their simulations, the wave diffusivity is given by $u_w^2/\tau$ with $\tau=1\,{\rm s}$.
Despite the similar simulations and results to \citet{Rogers2017}, the diffusivities for the $3M_\odot$ zero-age main sequence model ranges from $10^4\,{\rm cm}^2/{\rm s}$ near the convection zone to $10^6\, {\rm cm}^2/{\rm s}$ near the top of the simulation ($0.9R_\star$).
This may be because the simulations analyzed in \citep{Varghese2023} have lower diffusivities than the simulations of \citep{Rogers2017} (private communications, Varghese, 2023).

\citet{Higl2021} ran similar simulations to \citet{Rogers2017}.
They ran 2D pseudo-incompressible simulations in cylindrical geometry of a $3M_\odot$ zero-age main sequence star, but find convective velocities $\sim 10$ times larger than \mbox{\citet{Rogers2017}}.
They noticed there was significant chemical mixing outside of their convection zone, so measured the diffusivity of passive tracer particles, using the same algorithm as \citet{Rogers2017}.
As the mixing occurs outside the convection zone, they interpret this diffusivity as being due to internal gravity waves.
However, they find the diffusivity decreases steadily as they increase their simulation resolution, meaning they can only place upper bounds on mixing from internal gravity waves.
Their highest-resolution simulation places an upper bound $D_w<10^7\,{\rm cm}^2/{\rm s}$ across most of the radiative zone.
\mbox{\citet{Rogers2013}} and \citet{Varghese2023} find that $v_w$ is proportional to the convective velocity; since \mbox{\citet{Higl2021}} has convective velocities $\sim$$10$ times larger than \mbox{\citet{Rogers2013}} and \mbox{\citet{Varghese2023},} their upward bounds would correspond to $D_w<10^5\, {\rm cm}^2/{\rm s}$ for the more realistic convective velocities of \mbox{\citet{Rogers2013}} and \citet{Varghese2023}.
This is much smaller than the  diffusivities measured in \citet{Rogers2013}, though consistent with \mbox{\citet{Varghese2023},} as they find $D_w\sim 10^4\, {\rm cm}^2/{\rm s}$ in the inner part of the star, and \mbox{\citet{Higl2021}} only simulate to $0.5R_\star$.
Nevertheless, \citet{Higl2021} illustrates the need to study how chemical mixing from waves depends on the resolution of the underlying wave simulation.

\citet{herwig2023a} measure the chemical diffusivity outside of their convection zone, and attribute it to mixing from internal gravity waves.
They specifically invoke the mechanism of shear-induced mixing \citep{Zahn1992}, in which the diffusivity scales like the vorticity squared, or like the 1/$Ri$, where $Ri$ is the Richardson number.
They find their measured diffusivities scale like $1/Ri$ for low luminosity boosting factors, but like $1/Ri^2$ for higher luminosity boosting factors.
Extrapolating to higher resolution and realistic stellar luminosities, they estimate $D_w\sim 10^4\,{\rm cm}^2/{\rm s}$ near the radiative-convective boundary for their $25M_\odot$ stellar~model.

\subsubsection{Angular Momentum Transport by Waves}

Gravity waves represent a \textit{non-local} form of angular momentum transport in a star.
When waves are excited by convection, angular momentum is transferred from the convection into the waves.
The waves deposit this angular momentum when they damp, which can be far from the convection zone.
The angular momentum transport by gravity waves is known to be an anti-diffusive process which can spontaneously generate differential rotation.
This is observed in the Earth's atmosphere where convectively excited gravity waves drive the quasi-biennial oscillation (QBO; \citep{Baldwin2001}).

The 2D simulations of \citet{Rogers2013} exhibit the anti-diffusive nature of wave-driven angular momentum transport.
Even though their simulations start with solid-body rotation, in some simulations they find the surface layers begin to spin up due to wave-driven angular momentum transport.
The surface layers spin up first because the wave amplitude is largest near the surface because the density is low and waves propagate with roughly constant luminosity (Equation~(\ref{eqn:wave luminosity})).
After the surface spins up, subsequent waves which reach the surface encounter a critical layer, causing them to damp and deposit their angular momentum.
The entire radiative zone begins to spin up from the outside in.
This is very similar to one phase of the QBO.
Some simulations develop prograde rotation while others develop retrograde rotation.
They hypothesize the differential rotation may undergo oscillations on timescales longer than the simulation time.

\citet{Rogers2015} ran a series of 2D cylindrical anelastic simulations similar to those in \citet{Rogers2013}, but varied the luminosity of the model and their initial rotation rate.
She found that simulations with small luminosities could develop strong differential rotation between their core and envelope.
Simulations with higher luminosities also produced differential rotation, but with lower magnitude.
One of the simulations produced a differential rotation profile similar to the profile inferred from asteroseismic modeling of KIC 10526294 \citep{Triana2015}.
The simulations of \citet{Rogers2015} all have luminosities boosted by $\gtrsim$$10^3$, so they likely exhibit more efficient angular momentum transport than real stars.
As Rogers found different results for different luminosities, this work highlights the importance of estimating angular momentum transport by waves excited by convection carrying realistic luminosities.

\subsection{Spectra of Convection}
The common theoretical expectation of the spectra of kinetic energy in core convection is that of isotropic turbulence. That means the spatial spectrum should scale with $k^{-5/3}$ over a large range of $k$, called the inertial subrange, with $k$ being the magnitude of the three-dimensional wavenumber. In their 3D simulations \citet{Gilet2013} see good agreement with this prediction in a range from $k=10$ to $k=100$ for the kinetic energy spectrum averaged over the whole convection zone. This also holds for individual components in the three spatial directions, with a slight discrepancy between the two angular directions at low wavenumbers only, which the authors attribute to the overall flow morphology at large length scales.

The anelastic simulations of \citet{Edelmann2019} show a steeper spectrum with $k^{-2}$ to $k^{-3}$. They hypothetize that this could be evidence of Bolgiano--Obukhov scaling, but defer this to more detailed simulations of the core itself, which was not the focus of that work. \citet{horst2020a} used the same background state as \citet{Edelmann2019} but ran fully compressible 2D simulations using a lower luminosity. They see a scaling of $k^{-2.1}$ at the top of the convection zone.

\citet{herwig2023a} study the radial dependence of the kinetic energy within the convection zone. They find an almost perfect $k^{-5/3}$ spectrum deep in the convection zone but this becomes much flatter towards the convective--radiative interface. The explanation is that there is less space for large scale radial motions in the remaining distance to the interface, making the small scale motions relatively more important. This is supported by the fact that the spectrum based on horizontal velocities only is not subject to this effect.

 \section{Perspectives}

Core convection plays an important role in the structure and evolution of intermediate- and high-mass stars, but there remain many uncertainties in how to parameterize this intrinsically multidimensional process in one-dimensional stellar evolution models.
In this review, we have described recent efforts to run multidimensional simulations of core convection.
This is very challenging, as there are a wide range of temporal scales, spatial scales, and physical efforts that are important to capture.
Two of the largest issues are the short sound crossing time scale, and the long thermal time scale.
As described in Section~\ref{sec:challenges}, the sound time scale is $\sim$$10^4$ faster than the convection time, while the thermal time scale is $\sim$$10^6$ slower than the convection time, for typical stars of interest.

There are multiple strategies for addressing fast sound waves (equivalently, the low Mach number of convective flows); see Section \ref{sec:approaches} for more details.
One is to solve the full Navier--Stokes equations as usual, using explicit timestepping methods (e.g., PPMstar, PROMPI).
The CFL criterion requires such a simulation to take timesteps of size $\sim \Delta x/c_s$ where $\Delta x$ is the grid spacing and $c_s$ is the sound speed.
These codes require lots of timesteps, but use fast algorithms that can still make them efficient.
Another strategy is to use implicit timestepping algorithms (e.g., MUSIC, SLH, Dedalus).
These algorithms allow simulations to take timesteps of size $\sim \Delta x/u_c$, where $u_c$ is the convective velocity.
While implicit methods are more computationally expensive per timestep than explicit methods, each timestep can be larger by $c_s/u_c\sim 10^4$.
Finally, one can solve a different set of underlying equations that do not admit sound waves, e.g., the anelastic (e.g., \citep{Ogura1962}) or the pseudo-incompressible equations \citep{Durran1989}.
This approach is adopted by, e.g., MAESTRO, the Rogers group, ASH, Rayleigh, and allows one to take large timesteps of size $\sim \Delta x/u_c$ without requiring implicit timestepping of sound waves.
However, these equations involve solving an elliptic equation which increases the computational cost, as well as algorithmic complexity, of each timestep.

Another common strategy is to artificially increase the luminosity of the star.
The convective velocities scale like luminosity to the one third power, so increasing the luminosity by a factor of $10^3$ increases the convective velocity by a factor of $10$.
That would then decrease the convective time by a factor of $10$, making the ratio of convective to sound time scales closer to $10^3$.
This approach is used both by codes using explicit timestepping (e.g., PPMstar), as well as some which use implicit timestepping (e.g., some configurations of the SLH code \citet{horst2020a}).
However, great care must be taken when comparing these simulation results to real stars, \citet{Baraffe2021} and \citet{LeSaux2022} have shown that changing the luminosity has significant and complex effects on both convective boundary mixing and convective excitation of waves (though \citet{Anders2022penetration} finds convective penetration is roughly independent of luminosity).

Another major challenge in modeling core convection is the very long thermal time.
This is particularly important for understanding convective boundary mixing.
Entrainment rates which match stellar observations are orders of magnitude smaller than those measured in simulations \citep{Staritsin2013}.
This indicates that simulations are not initialized near thermal equilibrium, so these high entrainment rates are transients that may not be relevant in stellar evolution (which is slow relative to the thermal time).
One possible thermal equilibrium is an extended adiabatic layer beyond the Schwarzschild boundary, which is known as convective penetration (see Chapter 5).
As simulations cannot be run for a thermal time, new strategies must be developed for finding these thermal equilibria.
This includes increasing the stellar luminosity (which decreases the thermal time), as well as the ``accelerated evolution'' technique employed in \citet{Anders2022penetration}.

It is also important to simulate the high turbulent intensity of core convection.
Turbulence acts differently in two- and three-dimensional simulations, and modeling realistic flows requires three dimensions.
Furthermore, one must ensure there is sufficient scale separation between the radius of the convective core, which represents the size of large-scale fluid motions, and the dissipation scale of the simulation.
While no simulation will achieve the very small dissipation scales of real stars, there is great effort to decrease the dissipation scale as much as possible.
Luckily, it appears that current simulations have sufficiently small dissipation scales that the large-scale convective dynamics are insensitive to the dissipation scale.
However, other effects, such as wave mixing, may remain strongly resolution dependent \citep{Higl2021}.

The simulations presented in this review have yielded exciting preliminary results on core convection.
While many works analyze simulations with boosted luminosities and/or run in two dimensions, the numerical tools and computational resources now exist to run three-dimensional simulations with realistic stellar luminosities.
Going forward, these more realistic simulations will be crucial for providing more insights into the processes occurring in stars.

While some aspects of core convection, including kinetic energy spectra and average convective velocities, are similar across comparable simulations, there are discrepancies in other aspects.
Many groups have run simulations with different algorithms and analyzes, and have arrived at different answers to important questions regarding convective boundary mixing and wave mixing.
In fact, sometimes groups using similar codes and similar analyzes also find very different results.
To produce robust parameterizations for use in one-dimensional stellar modeling, different computational groups must work together to understand the differences between their simulations to determine what occurs in real stars.
This will require inter-group comparisons, such as \citet{Andrassy2022}, which established consistent convective entrainment rates across a range of codes in an idealized convection problem with Mach number $0.04$.
Future code comparison projects focused on convective boundary mixing and convective wave excitation will be able to provide accurate parameterizations of these multidimensional phenomena for stellar evolution~modeling.

Even though hydrodynamics simulations of convection can greatly improve our understanding of the physical processes occurring in stars, their overall evolution can only be studied using hydrostatic stellar evolution codes, simply due to the disparity of the evolutionary and dynamical timescales. That is why it is important to use the insights gained in hydrodynamic simulations and use them in stellar evolution codes (e.g.,~\citep{Battino2016,Battino2019,scott2021a,baraffe2023a}).

\vspace{6pt}

\authorcontributions{D.L. and P.V.F.E. conceptualized, wrote, reviewed, and edited this review article. All authors have read and agreed to the published version of the manuscript.}

\funding{D.L. is supported in part by NASA HTMS grant 80NSSC20K1280 and NASA OSTFL grant 80NSSC22K1738. P.V.F.E. was supported by the U.S. Department of Energy through the Los Alamos National Laboratory (LANL). LANL is operated by Triad National Security, LLC, for the National Nuclear Security Administration of the U.S. Department of Energy (Contract No. 89233218CNA000001).}

\acknowledgments{This work has been assigned a document release number LA-UR-23-23758.}

\conflictsofinterest{The authors declare no conflicts of interest.}

\begin{adjustwidth}{-\extralength}{0cm}

\reftitle{References}

\bibliography{references}

\PublishersNote{}
\end{adjustwidth}
\end{document}